\documentclass[fleqn,usenatbib]{mnras}

\usepackage{newtxtext,newtxmath}

\usepackage[T1]{fontenc}

\DeclareRobustCommand{\VAN}[3]{#2}
\let\VANthebibliography\thebibliography
\def\thebibliography{\DeclareRobustCommand{\VAN}[3]{##3}\VANthebibliography}

\usepackage{graphicx}	%
\usepackage{amsmath}	%
\usepackage{xspace}     %
\usepackage{subfig}     %
\usepackage{fp}
\usepackage{orcidlink} %
\usepackage{cellspace} %
\usepackage{makecell} %

\newcommand{\msol}{\mbox{M$_\odot$}}
\newcommand{\lsol}{\mbox{L$_\odot$}}

\newcommand{\yr}{\mbox{${\rm yr}$}}

\newcommand{\kms}{\mbox{${\rm km~s^{-1}}$}}

\newcommand{\cii}{\mbox{[C{\sc ii}]}}
\newcommand{\cplus}{\mbox{$\rm C^{+}$}}

\newcommand{\zcii}{7.30651}
\FPeval{\zciitwodp}{round(\zcii:2)} %
\FPeval{\zciifourdp}{round(\zcii:4)}
\newcommand{\zciierr}{0.0001}

\newcommand{\fluxcont}{\mbox{$260 \pm 22$}}
\newcommand{\lir}{\mbox{$1.5^{+0.8}_{-0.5} \times 10^{12}$}}

\newcommand{\sfrtot}{\mbox{$199^{+101}_{-63}$}}
\newcommand{\sfruv}{\mbox{$14 \pm 3$}}
\newcommand{\sfrir}{{\mbox{$ 185^{+101}_{-63}$}}}

\newcommand{\lciival}{\mbox{$1.7 \pm 0.2 \times 10^{9} $}}

\newcommand{\mmolcii}{\mbox{$5.1^{+5.1}_{-2.6} \times 10^{10}$}} %
\newcommand{\tdepl}{\mbox{$0.3^{+0.3}_{-0.2} $}}
\newcommand{\ciiinc}{\mbox{$56 \pm 29 $}}

\newcommand{\offsetsnr}{\mbox{$4.7$}}
\newcommand{\lciimerger}{\mbox{$3 \pm 1 \times 10^{8}$}}
\newcommand{\mmolmerger}{{\mbox{$9^{+9}_{-5}  \times 10^{9}$}}}
\newcommand{\sfrcii}{\mbox{$246 \pm 35$}}
\newcommand{\moutdproj}{\mbox{$200$}}
\newcommand{\mstar}{\mbox{$ 8^{+4}_{-2}  \times 10^{9} $}}
\newcommand{\nonparmass}{\mbox{$ 19^{+5}_{-8}  \times 10^{9} $}}
\newcommand{\sfrms}{\mbox{$ 49 $}}
\newcommand{\sfrmsnonpar}{\mbox{$ 101 $}}

\newcommand{\citetschouws}{\mbox{Schouws et al. (in prep.)}}
\newcommand{\citepschouws}{\mbox{(Schouws et al. in prep.)}}

\newcommand{\citetstefanon}{\mbox{Stefanon et al. (in prep.)}}
\newcommand{\citepstefanon}{\mbox{(Stefanon et al. in prep.)}}

\title[Discovery of a massive, complex ULIRG at $z =\zciitwodp{}$]{The ALMA REBELS Survey: Discovery of a massive, highly star-forming and morphologically complex ULIRG at $z =\zciitwodp{}$ }

\author[A. P. S. Hygate et al.]{A. P. S. Hygate \orcidlink{0000-0002-6488-471X},$^{1}$\thanks{E-mail: hygate@strw.leidenuniv.nl (APSH)}
J.~A.~Hodge \orcidlink{0000-0001-6586-8845},$^{1}$
E.~da~Cunha \orcidlink{0000-0001-9759-4797},$^{2,3}$
M.~Rybak \orcidlink{0000-0002-1383-0746},$^{4,1}$
S.~Schouws \orcidlink{0000-0001-9746-0924},$^{1}$
H.~Inami \orcidlink{0000-0003-4268-0393},$^{5}$
\newauthor
M.~Stefanon \orcidlink{0000-0001-7768-5309},$^{1,6}$
L.~Graziani \orcidlink{0000-0002-9231-1505},$^{7,8}$
R.~Schneider \orcidlink{0000-0001-9317-2888},$^{7,9,10,11}$
P.~Dayal \orcidlink{0000-0001-8460-1564},$^{21}$
R.~J.~Bouwens \orcidlink{0000-0002-4989-2471},$^{1}$
\newauthor
R.~Smit \orcidlink{0000-0001-8034-7802},$^{11}$
R.~A.~A.~Bowler \orcidlink{0000-0003-3917-1678},$^{14}$
R.~Endsley \orcidlink{0000-0003-4564-2771},$^{15}$
V.~Gonzalez \orcidlink{0000-0002-3120-0510},$^{16,17}$
P.~A.~Oesch \orcidlink{0000-0001-5851-6649},$^{18,19}$
\newauthor
D.~P.~Stark \orcidlink{0000-0001-6106-5172},$^{15}$
H.~S.~B.~Algera \orcidlink{0000-0002-4205-9567},$^{5,20}$
M.~Aravena \orcidlink{0000-0002-6290-3198},$^{21}$
L.~Barrufet \orcidlink{0000-0003-1641-6185},$^{18}$
A.~Ferrara \orcidlink{0000-0002-9400-7312},$^{22}$
\newauthor
Y.~Fudamoto \orcidlink{0000-0001-7440-8832},$^{23,20}$
J.~H.~A.~Hilhorst \orcidlink{0000-0001-5919-2295},$^{1}$
I.~De~Looze \orcidlink{0000-0002-7129-5761},$^{24,25}$
T.~Nanayakkara \orcidlink{0000-0003-2804-0648},$^{26}$ and
A.~Pallottini\orcidlink{0000-0002-7129-5761 },$^{22}$
\newauthor
D.~A.~Riechers \orcidlink{0000-0001-9585-1462},$^{27}$
L.~Sommovigo \orcidlink{0000-0002-2906-2200},$^{22}$
M.~W.~Topping \orcidlink{0000-0001-8426-1141},$^{15}$ and
P.~van~der~Werf  \orcidlink{0000-0001-5434-5942}$^{1}$
\\
\\
Affiliations are listed at the end of the paper
}

\date{Accepted XXX. Received YYY; in original form ZZZ}

\pubyear{2022}

\begin{document}
\label{firstpage}
\pagerange{\pageref{firstpage}--\pageref{lastpage}}
\maketitle

\begin{abstract}
We present Atacama Large Millimeter/Submillimeter Array (ALMA)  \cii{} and $\sim158$ $\rm \mu m$ continuum observations of REBELS-25, a massive, morphologically complex ultra-luminous infrared galaxy (ULIRG;  $L_{\rm IR} = $ \lir{} \lsol{}) at $z = \zciitwodp$, spectroscopically confirmed by the Reionization Era Bright Emission Line Survey (REBELS) ALMA Large Programme. REBELS-25 has a significant stellar mass of $M_{*} = \mstar{}  ~\msol{}$. From dust-continuum and ultraviolet observations, we determine a total obscured + unobscured star formation rate of SFR $ = \sfrtot{} ~ \msol ~ \yr^{-1}$. This is about four times the SFR estimated from an extrapolated main-sequence. We also infer a \cii{}-based molecular gas mass of $M_{{\rm H}_{2}} = \mmolcii{} ~\msol{}$, implying a molecular gas depletion time of $  t_{\rm depl, {\rm H}_{2}} =  \tdepl{}$ Gyr. We observe a \cii{} velocity gradient consistent with disc rotation, but given the current resolution we cannot rule out a more complex velocity structure such as a merger. The spectrum exhibits excess \cii{} emission at large positive velocities ($\sim 500$ \kms{}), which we interpret as either a merging companion or an outflow. In the outflow scenario, we derive a lower limit of the mass outflow rate of \moutdproj{}~$\msol ~ \yr^{-1}$, which is consistent with expectations for a star formation-driven outflow. Given its large stellar mass, SFR and molecular gas reservoir $\sim700$ Myr after the Big Bang, we explore the future evolution of REBELS-25. Considering a simple, conservative model assuming an exponentially declining star formation history, constant star formation efficiency, and no additional gas inflow, we find that REBELS-25 has the potential to evolve into a galaxy consistent with the properties of high-mass quiescent galaxies recently observed at $z \sim 4$.

\end{abstract}


\section{Introduction}
\label{sec:introduction}

A key frontier of astrophysics is understanding the emergence of the first galaxies: the transition from the Cosmic ``Dark Ages", through the Epoch of Reionization (EoR) at $6 < z < 11$ \citep{PlanckCollaboration2016} to the large variety of galaxy populations observed today. Thanks to extensive optical and near-IR surveys with the \textit{Hubble Space Telescope (HST)} and large ground-based telescopes, the observational study of galaxies has steadily been pushing to higher and higher redshift. With the recent observation of a galaxy potentially at $z \sim 11$ \citep{Oesch2014,Oesch2016,Jiang2021}, the study of galaxies has now been extended significantly towards the highest redshifts. Such studies, along with the discovery of other particularly massive (M$_*$ $\ge$ 10$^{9}$--10$^{10}$ \msol{}) galaxies at $z > 6$ \citep[e.g.][]{Riechers2013,Strandet2017,Marrone2018,Banados2019,Spilker2022} support the presence of significant early star formation and mass build-up in the first few hundred Myr after the Big Bang.

Due to the observational challenge presented by early galaxies, our understanding of these objects is rapidly evolving. Observations of galaxies in this era suggest that they are generally compact, and dust- and metal-poor, though significant variation exists \citep[see reviews by][]{Stark2016,DayalFerrara2018}. The majority of star formation during this era  has long been thought to be unobscured by dust, and well traceable by ultraviolet (UV) light, although the advent of the Atacama Large Millimeter Array (ALMA) is now pushing studies of dust-obscured star formation to the highest redshifts (see \citealt{HodgedaCunha2020} for a review), including the recent discovery of normal, dust-obscured galaxies at $z > 6$ with a high fraction of obscured star formation \citep{Watson2015,Fudamoto2021,Bakx2021}. Additionally, recent determinations of the star formation rate surface density from large surveys indicate that the contribution of obscured star formation may have been underestimated at high redshifts	\citep{Gruppioni2020,Talia2021,Viero2022,Algera2023a}. Indeed, a recent analysis of the history of the infrared (IR) luminosity function \citep{Zavala2021} suggests that obscured star formation is dominated by so-called ultra-luminous infrared galaxies with $L_{\rm IR} > 10^{12} \lsol$ \citep[ULIRGs; e.g.][]{Lonsdale2006} above $z\sim 2$. Such sources have, however, proven difficult to find in the EoR, with classical submillimetre surveys only sensitive to the most extreme examples at redshifts $z > 6$ \citep[e.g][]{Strandet2017,Marrone2018}.

At the same time, recent observations have revealed the presence of a population of massive (in excess of $\log{\rm M_{*}} /  \msol{} = 10.5$), high-redshift quiescent galaxies \citep[e.g.][]{Glazebrook2017,Schreiber2018a,Merlin2019,Carnall2020,Forrest2020,Forrest2020a,Saracco2020,Tanaka2019,Valentino2020,Kubo2021}. These passive (i.e. with star formation rate (SFR) significantly below the main sequence) galaxies have been confirmed spectroscopically out to $z\sim 4$, with photometric evidence suggesting an even higher-redshift population. The discovery of such a population naturally raises questions about how they formed and quenched on such short timescales, with some studies suggesting that their existence is a challenge to the latest hydrodynamical simulations and semi-analytical models \citep[e.g][]{Steinhardt2016,Schreiber2018a,Cecchi2019,Girelli2019}. Studies have shown that the classical sub-millimetre-selected galaxies (SMGs) at $z\gtrsim 4$ are plausible progenitors of these high-redshift quiescent galaxies \citep{Toft2014,Valentino2020}, though with increasing tension at the highest ($z > 5$) redshifts \citep{Valentino2020}. In particular, a detailed comparison by \citet{Valentino2020} of the number densities of high-redshift quiescent galaxies and SMGs shows broad agreement, although the current constraints on number densities of SMGs at $z > 5$ are limited \citep[see e.g.][]{Riechers2020a,Zavala2021}. Additionally, \citet{Valentino2020} find that some of the progenitors of high-redshift quiescent galaxies likely have less extreme progenitors than SMGs, with submillimetre fluxes not necessarily detectable in the deepest current submillimetre surveys. These progenitor galaxies would nevertheless have needed to build up significant stellar masses on short timescales through vigorous star formation.

An important and related question is how such highly star-forming galaxies at these high redshifts then transitioned via quenching from an active star formation phase to a passive phase. There are a number of mechanisms that can produce this effect, including outflows driven by star-formation and active galactic nuclei (AGN). Outflows at high redshift have been identified in absorption spectra, both in composite spectra made from galaxies at redshift \mbox{$3 < z < 7$} \citep{Jones2012} and spectra of individual lensed galaxies \citep{Jones2013}. Outflows have also been detected in a number of individual galaxies via OH$^+$ P-Cygni profiles \citep{Riechers2021a} and OH absorption \citep{Spilker2020a,Spilker2020}. Observed offsets between the Lyman-$\rm \alpha$ line and \cii{}, from which the galaxy's redshift can be established, are also used to infer the presence of outflows \citep{Cassata2020}. Evidence for high redshift outflows has also been slowly building from observations of \cii{} itself. Detections of outflows based on \cii{} have been reported in individual galaxies \citep{Cicone2015,Maiolino2012, Izumi2021} and stacking analyses \citep{Bischetti2019}, in particular \citet{Izumi2021} detected a galactic-scale outflow in a quasar at $z = 7.07$  (i.e. within the EoR), with a significant atomic mass outflow rate of $ \sim450 ~ \msol \yr^{-1} $ with a corresponding atomic mass loading factor of $ \sim3$ and an estimated total mass loading factor of $\sim9$. It remains an open question, however, how common outflows are in AGN host-galaxies. A stacking analysis by \citet{Bischetti2019} indicated outflows are a common feature in high-z AGN host-galaxies. However, other analyses do not find evidence that outflows are common \citep{Decarli2018,Novak2020}.

More generally, stacking of \cii{} emission from galaxies between $4 <z < 6 $ indicates the presence of star-formation driven outflows at these redshifts \citep{Gallerani2018,Galliano2018,Fujimoto2019a,Ginolfi2020}. Observations of OH absorption in a sample of 11 lensed dusty star forming galaxies within this redshift range detected outflows in 73\% of galaxies \citep{Spilker2020a}. While not an unbiased sample, this result suggests such outflows are indeed common in the high-redshift universe, but one can ask the question as to how significant a role they play in the quenching of star formation. Measurement of the atomic hydrogen mass loading factor of star formation driven outflows from stacking \citep{Ginolfi2020a} and a single $z\approx 5.5$ galaxy \citep{Herrera-Camus2021} indicate values below unity, less significant in general than those of AGN. This difference between the values of the mass loading factor at high-z mirrors the more well established difference between the mass loading factors measured in nearby galaxies, with star-forming (including starbursting) galaxies having mass loading factors usually around unity and  $\lesssim 4$, whereas AGN can have mass loading factors many times this \citep[see e.g.][]{Cicone2014,Fluetsch2019}.  However, the importance of AGN in driving outflows is unclear for these relatively low mass galaxies from theoretical studies \citep{Pizzati2020}. 

This paper uses data from the Reionization Era Bright Emission Line Survey (REBELS; \citealt{Bouwens2022}). REBELS is an ALMA large programme to identify and study massive galaxies with substantial interstellar medium (ISM) reservoirs in the EoR. Practically, the survey uses spectral-scanning to identify and observe the $\rm {}^{2}P_{3/2} \rightarrow  {}^{2}P_{1/2}$ transition line of singly ionised carbon (\cplus{})  at a wavelength of 157.7 $\rm \mu m$ (hereafter \cii{}), the $\rm {}^{3}P_{1} \rightarrow {}^{3}P_{0}$ transition of [O{\sc iii}] at a wavelength of 88.4 $\rm \mu m$ and the dust continuum in a sample of 40 UV-bright galaxies at $z >  6.5$.

Here we focus on one particular galaxy in the sample, REBELS-25, which is the strongest $\sim158$ \textmu m continuum emitter in the sample. REBELS-25 is located on the sky at a J2000 right ascension of $10^{\rm h}00^{\rm m}32.32^{\rm s}$ and declination of $+01\degr44\arcmin31\farcs3$, as determined from rest-UV imaging \citep{Stefanon2019}. REBELS-25's redshift has been established as \zciifourdp{} from a detection of the \cii{} line \citepschouws. Previous observations have indicated that REBELS-25 has a complex morphology. Analysis of HST rest-frame UV observations show three distinct UV sources identified by \citet{Stefanon2019}.\footnote{Note that REBELS-25 is referred to by \citet{Stefanon2019} and \citet{Schouws2022a} as UVISTA-Y3.}  Dust-continuum emission has also been detected close to, but offset from the position of these UV clumps \citep{Schouws2022a}. Thus, these UV-clumps may be regions of unobscured star-formation embedded in a dusty disc visible due to differential obscuration or they may indicate a merger. Based on the latest REBELS Large Program data, we find that REBELS-25 is the galaxy most robustly defined as a ULIRG\footnote{We note that such a definition of a galaxy population in terms of a single luminosity cut across all redshifts is necessarily somewhat arbitrary and indeed observational evidence indicates clear differences in the properties of ULIRG populations at different redshifts \citep[see e.g.][]{Rujopakarn2011}.} in the sample \citep[][and see Section 4.1]{Inami2022,Sommovigo2022}, thus uniquely enabling us to characterise a ULIRG in the EoR.

We present an analysis of new \cii{} and IR observations of REBELS-25 from the REBELS survey, in combination with existing IR and HST observations of the rest-frame UV. First, we present the data used in this analysis and previously determined properties of REBELS-25 in Section~\ref{sec:data} and we use this data to characterise REBELS-25, in particular the properties of its ISM, in Section~\ref{sec:methods}. We discuss the implications of these results and, in particular, whether REBELS-25 has the potential to evolve into a high-z quiescent galaxy, in Section~\ref{sec:discussion}. Lastly, we present our conclusions in Section~\ref{sec:conclusions}.

Throughout the paper, we adopt a standard $\rm \Lambda$CDM cosmology with values for the cosmological parameters of $H_{0} = 70 $km s$^{-1}$, $\Omega_\Lambda = 0.7$ and $\Omega_m = 0.3$. At a redshift of \zciitwodp, this translates to a luminosity distance of 72519 Mpc and a physical separation of 5.096 kpc for 1" on-sky. We also adopt a \citet{Chabrier2003} initial mass function (IMF) for a mass range of $0.1 - 300 \msol{}$.

\section{Data and source properties}
\label{sec:data}

Here we discuss the data used in this paper from ALMA and HST and a give a summary of a number of previously identified properties of REBELS-25.

\begin{figure}
	\centering
	\includegraphics[width=\columnwidth]{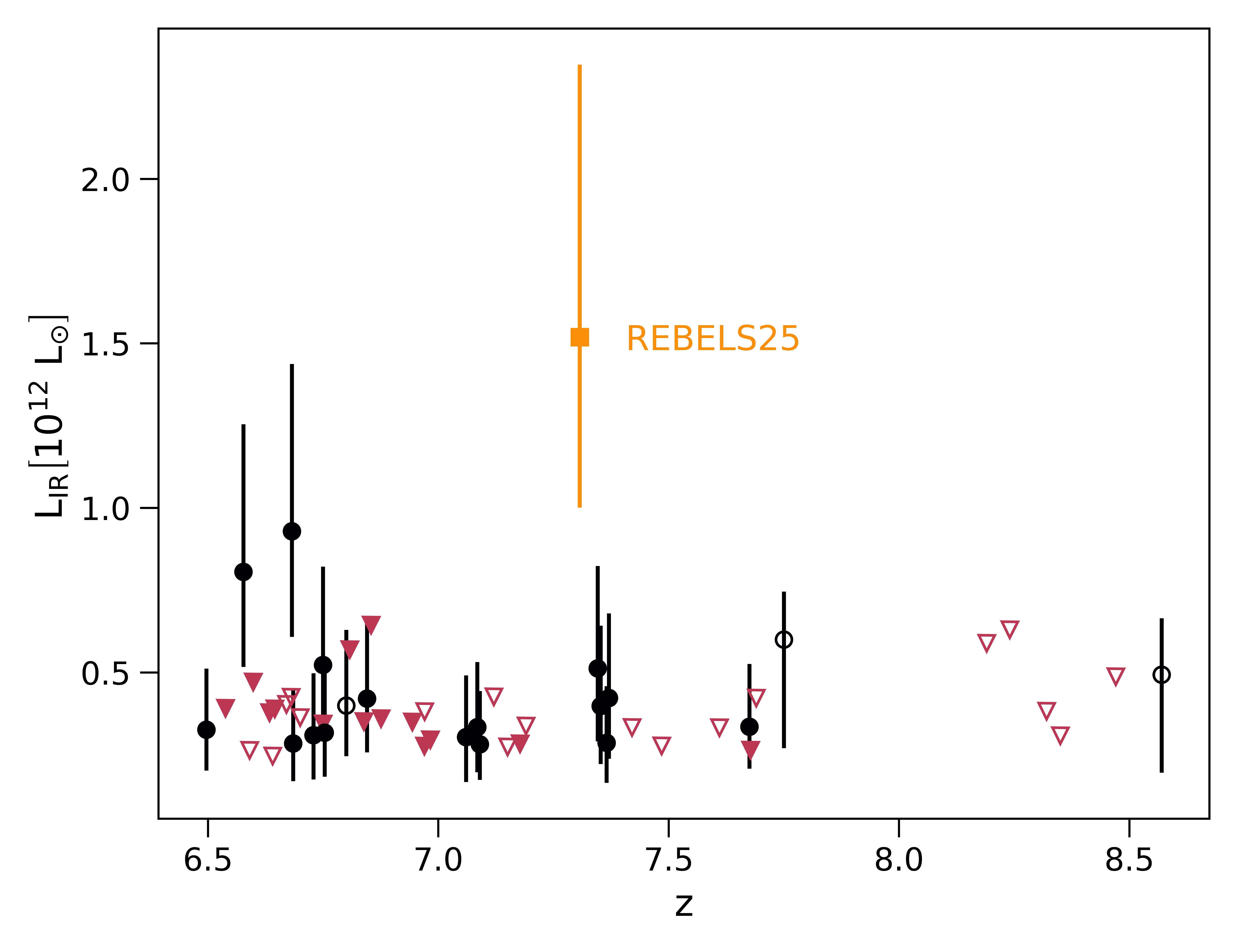}
	\caption{The infrared luminosities of galaxies in the REBELS survey \citep{Inami2022} including sources from the REBELS pilot programmes \citep{Smit2018,Schouws2022a,Schouws2022} against their redshifts. Sources with a measured infrared luminosity are shown as black circles with their 1-$\sigma$ errors and those where only a ($3\sigma$) upper limit on the infrared luminosity has been constrained are indicated with red downwards arrows. REBELS-25, the most infrared luminous source in the sample, is indicated with an orange square with 1-$\sigma$ error bars. Sources with a  \cii{}-determined redshift (presented in Schouws et al. in prep) are shown as filled symbols. The remaining sources are displayed as open symbols using their photometric redshifts \citep[presented in][]{Bouwens2022}. }

	\label{fig:rebels_sample}

\end{figure}

\subsection{ALMA Data}

We make use of ALMA C-1 and C-2 observations in Band 6 of the \cii{} line and dust continuum from the REBELS large programme (ID: 2019.1.01634.L). To trace the dust continuum, we also use additional higher resolution Band 6 C-4 observations (programme ID: 2017.1.01217.S). The full procedure employed in the reduction of these data is presented in \citetschouws{} and \citet{Inami2022} and we briefly summarise this process here. The data were reduced and calibrated with the ALMA Science Pipeline in version 5.6.1 of {\sc CASA}\footnote{The Common Astronomy Software Applications package, available at  \url{https://casa.nrao.edu/}} \citep{McMullin2007}. Subsequent analysis using {\sc CASA} was also performed with version 5.6.1. An image of the dust continuum was produced with the   {\sc CASA} command {\sc tclean}, excluding the region with identified \cii{} line emission.
To serve as the basis for \cii{} imaging, the calibrated measurement set was then continuum subtracted with a zeroth order polynomial fitted to the continuum, excluding a spectral region with a width of twice the FWHM of the identified line emission. The \cii{} line was then imaged using the {\sc tclean} task, by cleaning down to a 2$\sigma$ threshold using a clean mask generated with {\sc tclean}'s automasking feature, with natural weighting in order to maximise sensitivity to emission for use in determining integrated quantities. We also reimage the data with a Briggs robust weighting of 0.5 to obtain higher resolution imaging with which to investigate the morphology of REBELS-25. For the Naturally-weighted imaging, we obtain a beam of 1\farcs64 $\times$ 1\farcs32 for the \cii{} data cube and 1\farcs06 $\times$ 0\farcs94 for the continuum image. For the imaging produced with a robust weighting of 0.5, we obtain a beam of 1\farcs47  $\times$ 1\farcs15 for the \cii{} datacube and for the continuum image we achieve a beam of 0\farcs64 $\times$ 0\farcs56.

\subsection{HST data}
\label{sec:hst_data}

To trace the rest-frame UV emission, we use the COSMOS-DASH mosaic \citep{Mowla2019}. This is constructed by mosaicing together HST Wide Field Camera 3 (WFC3) F160W imaging from the COSMOS-DASH survey and the HST archive using the drift and shift (DASH) method presented by \citet{Momcheva2017}. The image reaches a depth of 25.1 mag (as calculated using a 0\farcs3 aperture; \citealt{Mowla2019}). The F160W filter has a pivot wavelength of 1543.17 nm\footnote{as listed in version 13.0 of the HST WFC3 Instrument Handbook, available at \url{https://hst-docs.stsci.edu/wfc3ihb}}. At the redshift of REBELS-25, this translates to a wavelength of 185.78 nm, i.e. rest-frame UV emission.

Astrometric offsets of HST imaging from that of ALMA are well-established \citep[see e.g.][]{Dunlop2017}. Thus, to improve the spatial comparability of the two datasets, we apply an astrometric correction to this image to align it with Gaia \citep{GaiaCollaboration2016}. We use the python tool {\sc astrometry} \citep{Wenzl2022} to calculate a correction to the astrometry of the image by comparing the positions of sources in GAIA Data Release 3 \citep[DR3][]{GaiaCollaboration2022} to their identified positions in our image. This correction relies on the astrometric solution for Gaia DR3, which is presented by \citet{Lindegren2021};  the celestial reference frame of Gaia DR3, which is presented by \citet{GaiaCollaboration2022a} and the assessment of the astrometric quality of Gaia DR3, which presented by \citet{Fabricius2021}. We estimate the remaining uncertainty in the astrometry of the image after we align it to Gaia to be $\sim20$ milliarcseconds, based on the rms of the distances between the catalogue positions of Gaia sources and their positions in our astrometrically corrected image.

\subsection{Source identification and properties}
\label{sec:source_properties}

\cii{} line detections for the REBELS sample are presented by \citetschouws{}. REBELS-25 is strongly detected in \cii{}, being one of the most \cii{}-luminous sources in the REBELS sample \citepschouws. The redshift of the sources are determined from the redshifted frequencies of the \cii{} line; for REBELS-25 this establishes the redshift as $\zciifourdp \pm \zciierr{}$ \citepschouws.  This spectroscopic redshift is in agreement with the photometric redshift of $z= 7.40^{+0.22}_{-0.19}$ determined for REBELS-25 by \citet{Bouwens2022} and lower than, but only slightly outside of the $1\sigma$ uncertainty of, the photometric redshift of $z =  7.62^{+0.14}_{-0.28}$ previously determined by \citet{Stefanon2019}.\footnote{We note that  \citet{Stefanon2019} determined different photometric redshifts depending  on whether they considered the UV emission to emanate from a single source or multiple sources. We discuss this further in Section~\ref{sec:morphology}.}  An independent study by \citet{Bowler2020} determined a photometric redshift for REBELS-25\footnote{\citet{Bowler2020} refer to REBELS-25 as UVISTA-213.} of $z =  7.39^{+0.12}_{-0.14}$ from fitting the UV continuum alone and  $z =  7.43^{+0.13}_{-0.16}$ from fitting the UV continuum and emission lines. Both of these photometric redshifts are consistent with the \cii{}-determined spectroscopic redshift to within their 1-$\sigma$ errors.

\citet{Inami2022} measure a 158~\textmu m  flux  of \fluxcont~\textmu Jy for REBELS-25. As the majority of the REBELS galaxies currently only have continuum observations at a single wavelength, their dust masses, dust temperatures and infrared luminosities cannot be constrained via SED fitting. The REBELS sample therefore employs the method presented in \citet{Sommovigo2021a}, which utilises the $158 $ \textmu m continuum in conjunction with the \cii{} luminosity to constrain these values. This method was applied to the thirteen REBELS galaxies detected in both \cii{} and continuum by  \citet{Sommovigo2022}. \citet{Sommovigo2022} determine a median dust temperature for the REBELS sample of $46^{+7}_{-6}$ K and present a conversion to infrared luminosity $L_{\rm IR}$, as $L_{\rm IR} = \alpha_{\rm IR} \nu_{\rm rest} L_{\nu_{\rm rest} }$, where $\nu_{\rm rest}$  is the rest frequency of the continuum observation and $\alpha_{\rm IR} = 14^{+8}_{-5}$ is the median conversion factor derived for the REBELS sample. This conversion is then applied uniformly to the whole REBELS sample in \citet{Inami2022}. We discuss the adoption of a dust temperature of 46K in context with the observational literature in Section~\ref{sec:ulirg}. In brief, the method employed by \citet{Inami2022} begins with the measurement of a monochromatic continuum flux density at rest-frame $\sim  158$\textmu m. This flux density is then corrected for the CMB contribution, under the assumption of a dust temperature of 46 K and converted to $L_{\rm IR}$ following the relation presented by  \citet{Sommovigo2022}. We show the IR luminosities determined by \citet{Inami2022} for the galaxies of the REBELS sample against their redshift in Figure~\ref{fig:rebels_sample}. This figure illustrates that, of the galaxies in the REBELS sample with IR luminosities presented by \citet{Inami2022}, REBELS-25 is the most IR luminous ($L_{\rm IR} = \lir{} \lsol{}$) and moreover it is the only galaxy whose luminosity qualifies it as a ULIRG.

REBELS-25 has a large stellar mass $M_{*} =  \mstar{}~\msol{}$ \citep[][Stefanon et al. in prep]{Bouwens2022}, which is measured by fitting the available rest-frame UV and optical photometry of REBELS-25 with the SED modelling code {\sc beagle} \citep{ChevallardCharlot2016}. In brief, {\sc beagle} self-consistently models a galaxy's SED using the \citet{BruzualCharlot2003} stellar population synthesis models and \citet{Gutkin2016} nebular emission models to fit the available photometry for the source. For REBELS-25,  a constant star formation history and a \citet{Chabrier2003} IMF were assumed.  For the full details of the modelling procedure, we refer the reader to \citetstefanon{}, where this is presented. An alternative method of determining stellar masses using a non-parametric star formation history was presented in \citet{Topping2022}, which results in a higher stellar mass of \nonparmass{}~\msol{} for REBELS-25. We note in our analysis how this would affect our results. 
	
In addition, REBELS-25's high IR luminosity implies both a significant build up of dust at an early time in the Universe's history and a correspondingly high amount of obscured star formation. The large molecular gas reservoir that can be inferred from the \cii{} luminosity (see Section~\ref{sec:ism_properties}) also implies that REBELS-25 will continue to experience significant growth into the future. All of these properties combine to make REBELS-25 a particularly interesting galaxy from amongst the REBELS sample.

\section{Methods and results}
\label{sec:methods}

In this section, we use the data presented in section~\ref{sec:data} to investigate the physical nature of REBELS-25. We summarise a number of the key properties that we derive in Table~\ref{tab:rebels25_properties}.

\begin{table*}
	\centering\setcellgapes{2pt}\makegapedcells
	\caption{Table showing the properties of REBELS-25 presented in this paper and those properties used in this paper from other sources.}
	\label{tab:rebels25_properties}
	\begin{tabular}{lcc} %
		\hline
		Property & Value & Further Details \\
		\hline
		$z$   & $ \zciifourdp  \pm \zciierr $ &\citetschouws \\
		$L_{\rm IR} $   &  \lir{}~\lsol{}  & \citet{Inami2022} \\
		$M_{*}$     & \mstar~\msol{} & \citet{Bouwens2022} \& \citetstefanon{} \\
		$L_{\cii{}}$  &  \lciival{}~\lsol{}  & Section~\ref{sec:spectral_fitting}\\
  		$ M_{\rm dyn} $       &  $2-8 \times 10^{10} \msol$ & Section~\ref{sec:dynamical mass} \\
		$M_{{\rm H}_{2},\cii{}}$  &  \mmolcii{}~\msol{} & Section~\ref{sec:ism_properties}\\
    	SFR$_{\rm tot}$ &  \sfrtot{}~\msol yr$^{-1}$ & Section~\ref{sec:star_formation} \\
        SFR$_{\cii{}}$       & \sfrcii{}~\msol yr$^{-1}$ & Section~\ref{sec:star_formation}\\
        $ t_{\rm depl, {\rm H}_{2}} $      &  \tdepl{}~Gyr &  Section~\ref{sec:star_formation} \\
		\hline
	\end{tabular}
\end{table*}

\subsection{Spectral analysis}
\label{sec:spectral_fitting}

\begin{figure*}
	\centering
	\includegraphics[width=0.8\textwidth]{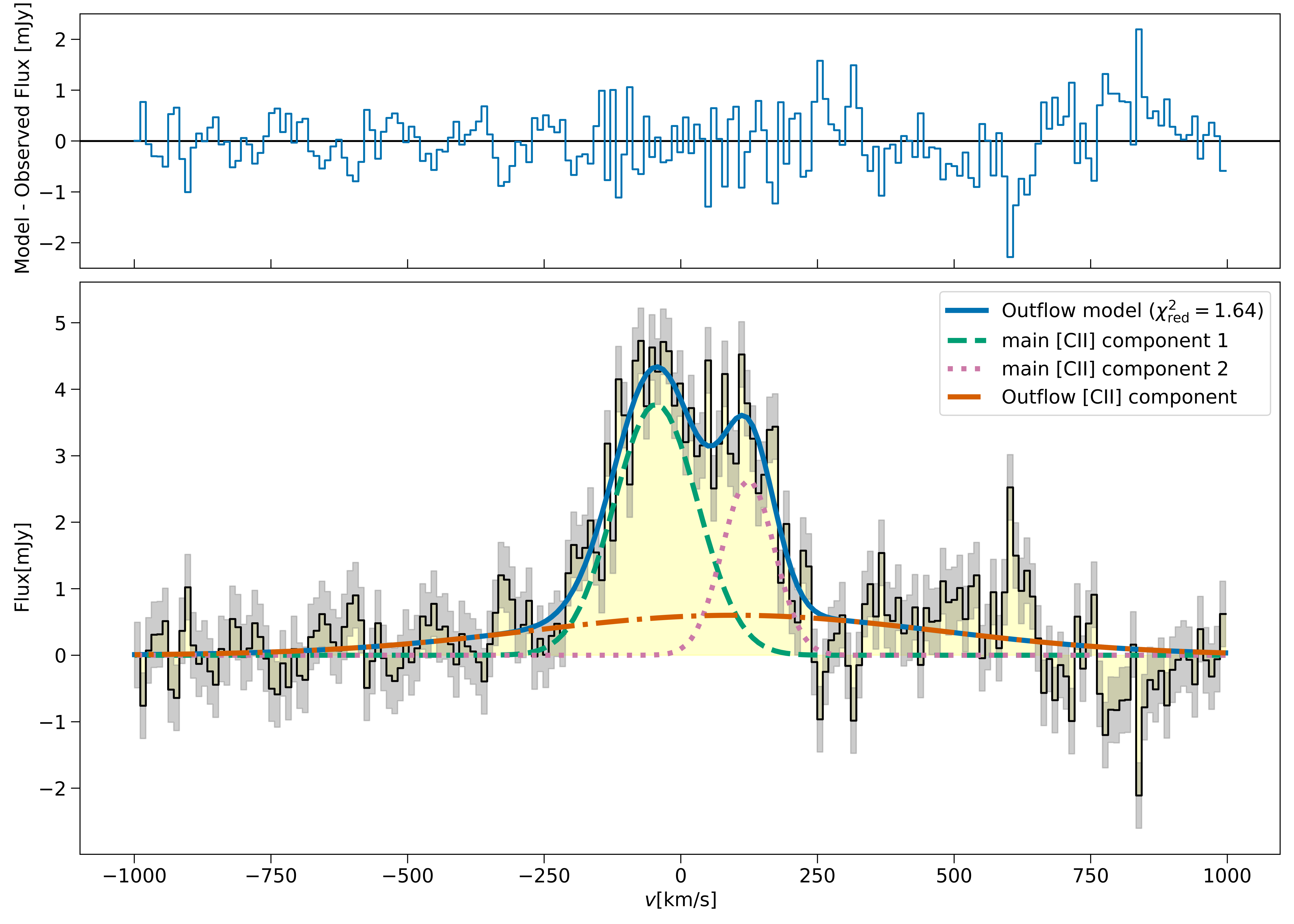} \\
	\includegraphics[width=0.8\textwidth]{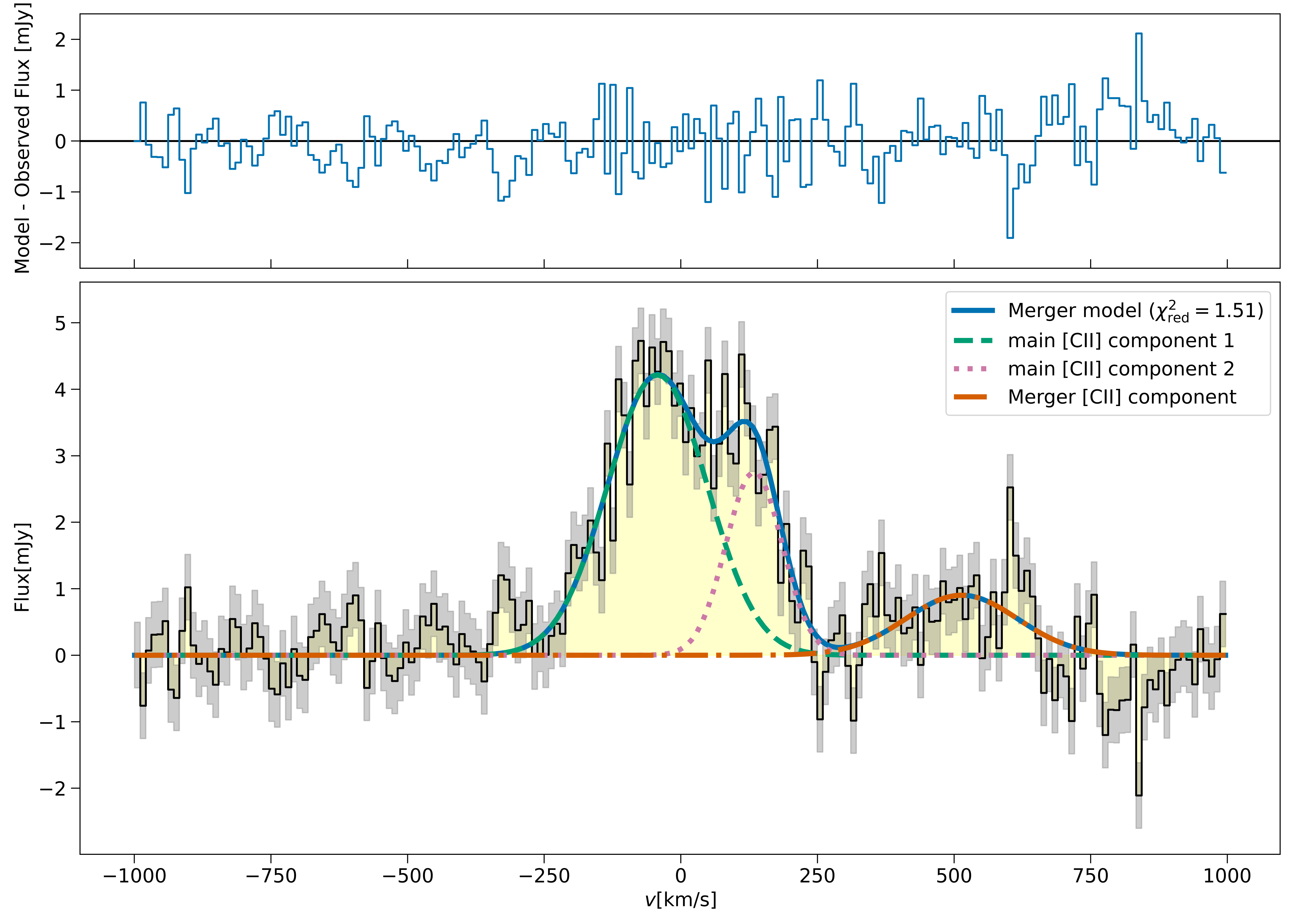}
	\caption{The continuum subtracted \cii{} spectrum of REBELS-25. The data itself is indicated with the black line and with yellow shading between the spectrum and the $y = 0$ line. The grey shaded area indicates the error on the data. The spectrum appears to show a double-peaked primary emission component and another fainter secondary component that is separated in frequency from the main component. We show the two best-fitting models to the data in the Top (``outflow'' model) and Bottom (``merger'' model) panels. For both models the blue solid line shows the model itself, the green dashed and purple dotted lines show the two Gaussian components fitted to the double-peaked primary emission component and the red dash-dotted line shows either a broad Gaussian, representing an outflow [Top panel] or a narrow Gaussian centred on the secondary emission component representing a merging galaxy [Bottom panel]. In the panels for each model, we display the residual spectrum (blue line, the difference between the model and observed flux) in inset panels above the spectrum itself. We discuss the details of the spectral analysis and models in Section~\ref{sec:spectral_fitting}.}
	\label{fig:cii_spectrum}
	
\end{figure*}

REBELS-25 is marginally resolved in \cii{} emission (see Section~\ref{sec:morphology}). As such, we extract a spectrum by placing a circular aperture centred at the position of the galaxy of radius 1\farcs75 ($\sim 9$ kpc), this radius being that at which the flux recovered from the galaxy plateaus. This \cii{} spectrum, shown in Figure~\ref{fig:cii_spectrum}, reveals a bright primary component (roughly in the range  -250 \kms{} to +250 \kms{}) that appears to be double-peaked as well as a fainter component that is evident as excess emission towards positive velocities (roughly in the range  +250 \kms{} to +650 \kms{}). We use the {\sc CASA} command {\sc immoments} to collapse the cube over the spectral range of these lines and create a zeroth moment map. From these, we measure a peak signal-to-noise ratio of 29 for the primary component and \offsetsnr{} for the secondary component. In order to confirm the presence of the lower signal to noise secondary component, we independently image two halves of our ALMA data in Appendix~\ref{sec:appendix_data_split} and find that it remains visible and has a signal to noise ratio in excess of three in both halves.

The faint secondary emission component could represent material outflowing from REBELS-25 or infalling into it, either in the form of an inflow or a merging galaxy. In this paper we consider the scenario of an outflow and that of a merging galaxy and note that while the specific evolutionary outcomes would be different, an inflow would lead to an increase in the amount of gas in REBELS-25, as with the merger scenario. Due to the range of possibilities for what the emission components could represent, we fit a range of models to the spectrum. We model the primary emission component with either one or two Gaussians. We are motivated to consider a model with two Gaussians by the observed double-peaked profile, but we also consider a model with only one Gaussian as a conservative alternative in order to assess the significance of the observed double peak. The observed double peak is likely indicative of unresolved velocity structure, most likely as a result of an unresolved major merger or disc rotation. We fit the secondary component with either a broad Gaussian to model an outflow or a narrow Gaussian to model a merging galaxy and as a control we fit models without any model component representing the secondary emission component.

We perform fits of our models using the {\sc optimize.curve\_fit} task in the {\sc scipy} python package\footnote{Available at \url{https://www.scipy.org/}}. We impose boundary conditions on the fit such that the Gaussians must have positive amplitudes (i.e. that they represent emission components and not absorption components). We also restrict the range in frequency space in which the different components can be centred. For the Gaussian(s) modelling the main emission component, we restrict them to being centred in the frequency range of the main emission component and thus not in the frequency range of the secondary emission component. For the Gaussian representing the secondary emission component, we require that its centre be within the frequency range of this secondary emission component in the case of the merger model or that it be within the frequency range of the main and secondary emission components for the outflow model. We additionally require that the outflow-representing broad Gaussian have a standard deviation in excess of 150 \kms{}, in order to distinguish it from the merger model.

We assess the quality of fit for each of the models by calculating the reduced $\chi^2$ parameter. We present all the fitted models and their reduced $\chi^2$ values in Table~\ref{tab:spectrum_models} in Appendix~\ref{sec:spectral_models}. Regardless of the model we chose for the secondary  component, a model with a double-peaked main component improves the quality of the fit. For the set of models with a double-peaked primary component, the two models that represent the secondary component as a merger or an outflow result in a better fit than the model without a component for this secondary emission.  The model with a double-peaked main component and merger (hereafter the merger model) is our best-fitting model with reduced $\chi^2 = 1.51$, but is statistically indistinguishable from the model with a double-peaked main component and outflow (hereafter the outflow model), which has reduced $\chi^2 = 1.64$. As the double-peaked main component most likely arises from an unresolved merger or a rotating disc, our best-fitting merger model would correspond to a triple merger system in the case where the main double-peaked emission component is a merger or a main rotating disc galaxy with a merging companion in the case where the main component is a rotating disc. The outflow model would then correspond to a merger with an outflow or a rotating disc with an outflow. We display both these models in Figure~\ref{fig:cii_spectrum} and we present the inferred properties of the secondary component, when considering it to be an outflow or a merging galaxy, in Section~\ref{sec:offset}. From the best-fitting merger model, we calculate a total \cii{} luminosity of \lciival{}~\lsol{} for the main \cii{} component from the two Gaussians with which we model it. We adopt this value for our main analysis, but note that calculating the luminosity from the statistically indistinguishable outflow model would instead lead to a reduced luminosity for the main component of $1.3 \pm 0.1 \times 10^{9} $ \lsol{}, due to some of the flux from the main \cii{} component being assigned to the outflow in this model.

\subsection{Morphology}

\label{sec:morphology}

\begin{figure}
	\centering
	\includegraphics[width=\columnwidth]{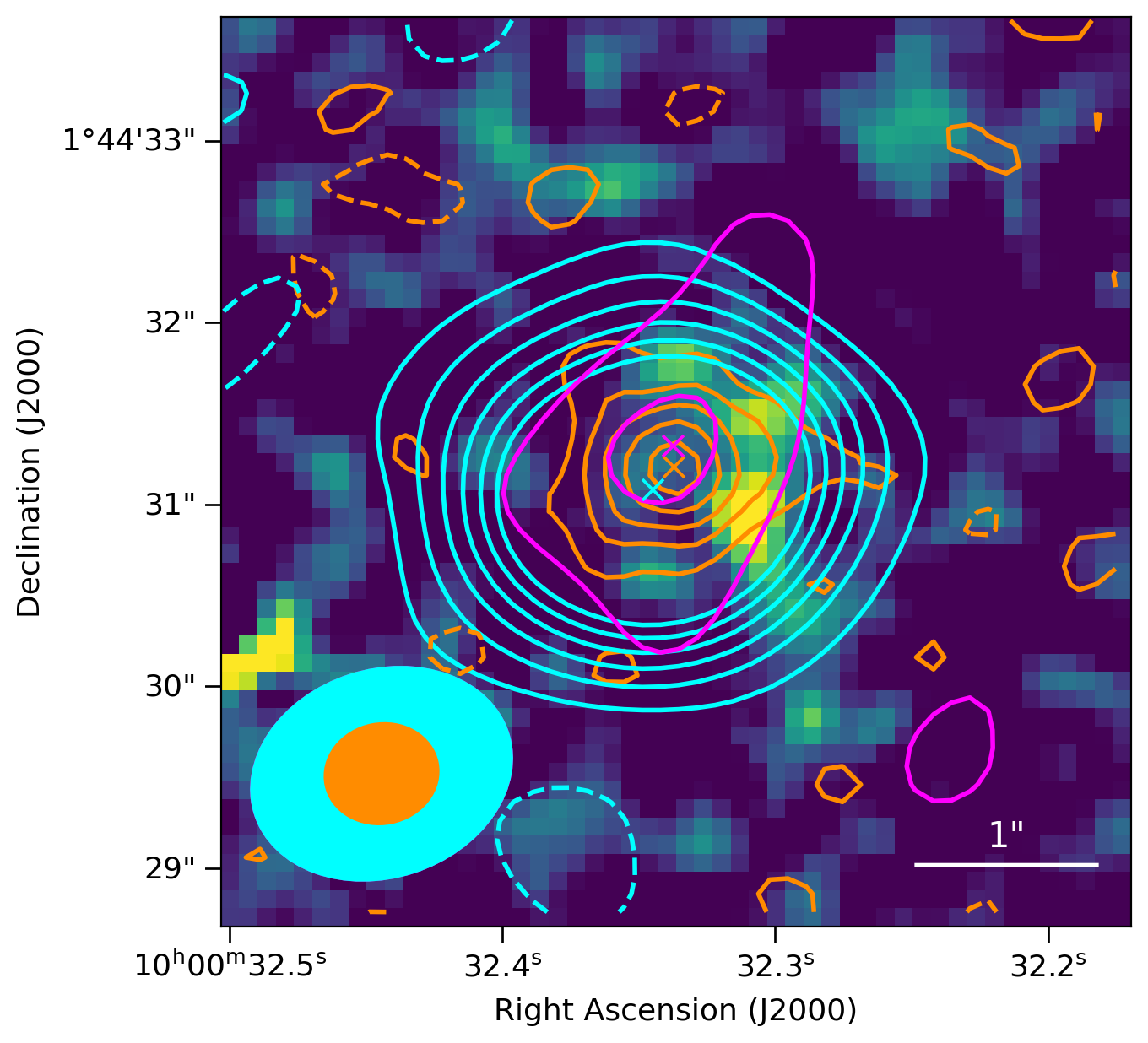}
	\caption{ A comparison of the morphologies of the emission observed at rest-frame UV and Far IR wavelengths from REBELS-25. A $5\arcsec \times 5\arcsec$ map of the HST COSMOS-DASH mosaic image \citep{Mowla2019} is displayed, which traces the rest-frame UV emission and shows the  three UV clumps identified by \citet{Stefanon2019}. We overlay our ALMA imaging of the dust continuum emission (orange contours), zeroth-moment of the main \cii{} emission component  (cyan contours) and the zeroth-moment of the fainter secondary \cii{} emission component (magenta contours). We show contours from $2\sigma$ up to a maximum of $12\sigma$, in steps of $2\sigma$ with solid contours. We also show negative emission with dashed contours at $-2\sigma$. We note that no negative emission from the secondary (magenta) zeroth-moment map is significant enough to be displayed within the field of view of the figure. A cyan cross marks the centre of the main \cii{} emission component, a magenta cross marks the centre of the secondary \cii{} component and an orange cross marks the centre of the dust continuum emission. The positions of these components, which are identified from image-plane fitting in {\sc CASA}, are close to  one another and appear to be offset from the positions of the UV emission components. 	All the displayed ALMA data is produced by imaging with a Briggs robust weighting parameter of 0.5. We display the resulting beams in the bottom left hand corner, for the continuum image as a solid orange ellipse and for both the \cii{} images as a solid cyan ellipse. A white scalebar showing one arcsecond on sky is displayed in the bottom right hand of the image.}

	\label{fig:mom0}

\end{figure}

We use the task {\sc immoments} in CASA to create a zeroth-moment map of the main and secondary \cii{} emission components, which we display along with the dust continuum and rest-frame UV HST imaging in Figure~\ref{fig:mom0}. In order to characterise the morphology of REBELS-25, we fit a Gaussian model to our images produced with a robust weighting of 0.5\footnote{We note that the values derived by applying the same analysis to the images produced with natural weighting are consistent within their errors.} using the task {\sc imfit} in CASA. We perform these fits for the main \cii{} emission component, the secondary \cii{} emission component and the continuum emission. From this fitting, we determine that the central positions of the main \cii{} component emission and the continuum are offset by one another by 0\farcs17 $\pm$~0\farcs04  (about 0.9 $\pm~0.2$ kpc) and that the secondary (spectrally-separated) \cii{} component is (spatially) offset from the main \cii{} component by about 0\farcs3 $\pm$ 0\farcs1 (about 1.5 $\pm~0.6$ kpc). However, we note that these uncertainties for the offsets only consider the uncertainties from the {\sc imfit} fitting process and do not take account of the uncertainty of the position of the sources within the ALMA image itself, which due to the low signal to noise is likely to be significant. Thus higher resolution observations are required to confirm these offsets.

The best-fitting Gaussian for the dust emission has deconvolved major and minor axis FWHMs of $0\farcs6~\pm~0\farcs2~\times~0\farcs5~\pm~0\farcs2$ ($ 3.1~\pm~0.8$ kpc  $\times~2.6~\pm~0.9$ kpc). We note that this is in good agreement with the size of $0\farcs74~\pm~0\farcs17~\times~0\farcs69~\pm~0\farcs22$ that was measured by \citet{Inami2022} using the lower resolution naturally-weighted imaging. The best-fitting Gaussian for the main \cii{} component has deconvolved major and minor axis FWHMs  of $0\farcs7~\pm~0\farcs1~\times~0\farcs4~\pm~0\farcs3$ ($ 3.6~\pm~0.7$ kpc  $\times~2.0~\pm~1.5$ kpc). Both the emission from the main \cii{} component and the dust emission have sizes consistent with each other to within the errors. Under the assumption of a thin circular disc, we then make a rough estimate of the inclination of the source using the ratio of the major and minor axes $i = \cos^{-1}{\theta_{min} / \theta_{max}}$. From the main component of the \cii{} and the continuum, we estimate inclinations of $ i = \ciiinc{} \degr{}$  and  $i = 31~\pm~43 \degr$, respectively. Lastly, the best-fitting Gaussian for the secondary \cii{} component has major and minor axis FWHMs $2\farcs2~\pm~0\farcs5~\times~1\farcs2~\pm~0\farcs2$ ($ 11.2~\pm~2.4$ kpc  $\times~6.0~\pm~0.8$ kpc).   However, we caution that the sources are only marginally resolved at the current angular resolution of the data.  The major and minor axes of the  convolved dust continuum source are resolved by 1.4 major beams and 1.6 minor beams by radius, respectively. The major and minor axes of the  convolved main \cii{} emission source are resolved by 1.1 major beams and 1.3 minor beams by radius, respectively.

Intriguingly, REBELS-25 is observed to have multiple separate components in the rest-frame UV \citep{Stefanon2019}. This could suggest a merger between multiple galaxies, or, as previously suggested for REBELS-25 by \citet{Ferrara2022}, these UV components could be clumps of star formation embedded within a dusty host galaxy and visible due to differential obscuration in different areas of the galaxy. \citet{Schouws2022a} considered the relative position of the rest-frame UV and dust continuum, using ALMA continuum data from project 2017.1.01217.S. They found that the centroid of the dust continuum emission was offset from, but in close proximity to the UV components. This remains the case when comparing the positions of the UV components to our imaging of the dust continuum that includes additional data from the REBELS large programme (see Figure~\ref{fig:mom0}). We compare the positions of the UV components identified by \citet{Stefanon2019} to the central position of the dust continuum emission, the main \cii{} component and the secondary \cii{} component. We find that the three UV components identified by \citet{Stefanon2019} are offset from the centre of the dust continuum by between $0\farcs5$ and $0\farcs7$ (2.6 - 3.3 kpc), the main  \cii{} component by between $0\farcs6$ and $0\farcs8$ (2.9 - 4.1 kpc) and the secondary \cii{} component by between $0\farcs5$ and $0\farcs6$ (2.3 - 3.1 kpc).

In order to assess the reliability of these offsets we consider the astrometric uncertainty of our ALMA imaging. The nominal uncertainty on astrometric positions of ALMA data is given by $\Delta\theta = {\rm FWHM}_{\rm beam} / SNR / 0.9$ \citep{Cortes2022}\footnote{Equation 10.7 in section 10.5.2 of the ALMA technical handbook, which is available from \url{https://almascience.eso.org/documents-and-tools/cycle9/alma-technical-handbook} }, where ${\rm FWHM}_{\rm beam}$ is the FWHM of the synthesised beam and SNR is the peak signal to noise ratio (however, improvement of the signal to noise ratio above 20 gives no improvement in the astrometric uncertainty). For our \cii{} observations we calculate a nominal uncertainty of $\sim 0\farcs1$. However, as the true astrometric uncertainty can be up to a factor of two worse than the nominal uncertainty \citep{Cortes2022}\footnote{see section 10.5.2 of the ALMA technical handbook, which is available from \url{https://almascience.eso.org/documents-and-tools/cycle9/alma-technical-handbook}}, we report an astrometric uncertainty of $\sim 0\farcs2$. For the dust continuum map, we are able to achieve higher resolution, but have a lower peak signal to noise. The combination of these two effects gives the continuum image an astrometric uncertainty, including the factor of two worsening, of  $\sim 0\farcs1$, about twice as good as our \cii{} imaging. We estimated a remaining uncertainty of $\sim$ 20 milliarcseconds after the alignment of the UV image to Gaia (see Section~\ref{sec:hst_data}), which is about 10 (20) per cent of the nominal ALMA uncertainty for the \cii{} (dust) beam. In addition, there can be an offset from ALMA's astrometric celestial frame  to other celestial reference frames of up to 23 milliarcseconds \citep{Cortes2022}\footnote{see section 10.5.2 of the ALMA technical handbook, which is available from \url{https://almascience.eso.org/documents-and-tools/cycle9/alma-technical-handbook}}, again this is about 10 (20) per cent of the nominal ALMA uncertainty for the \cii{} (dust) beam. Thus, for the main \cii{} component, the uncertainty on the offset to the UV position is dominated by the uncertainty in ALMA's astrometry. For the uncertainty in the offset to the dust continuum position these additional uncertainties are a more significant fraction of the better astrometric uncertainty we are able to achieve, but are still less significant than the astrometric uncertainty of our ALMA image. For the secondary \cii{} component there is also a significant contribution to the uncertainty from the measurement of the central position in {\sc imfit}, as the uncertainty on the right ascension co-ordinate is $\sim 0\farcs2$. Thus, we conclude that the the offsets that are observed between the UV components and both the dust and main \cii{} component centre  are larger than the uncertainties. In contrast, the offsets between the secondary \cii{} component and the UV components, the dust continuum centre and main \cii{} component centre are comparable to the uncertainties.

However, as we have no spectroscopic information on the clumps identified by \citet{Stefanon2019}, only broadband imaging, we note the additional caveat that we have assumed that the UV components are close to the observed \cii{} emission in velocity space. \citet{Stefanon2019} derived a photometric redshift for the UV components, assuming they represent a single source, of $z =  7.62^{+0.14}_{-0.28}$, consistent with our \cii{}-derived redshift to within 1.1 $\sigma$. \citet{Bowler2020} subsequently independently measured a UV photometric redshift for REBELS-25 of $z =  7.39^{+0.12}_{-0.14}$ from fitting to the UV continuum and $z =  7.43^{+0.13}_{-0.16}$ from fitting to UV emission lines in addition to the continuum. Both of these photometric redshifts are consistent with the  \cii{}-derived redshift. Most recently, \citet{Bouwens2022}  derived a photometric redshift of $z= 7.40^{+0.22}_{-0.19}$, which is consistent with our \cii{}-derived redshift of \zciitwodp{} \citep[][Schouws et al. in prep.]{Bouwens2022}. Treating them instead as separate sources, \citet{Stefanon2019} obtained photometric redshifts between $8.68$ and $9.29$, but with significant $1-\sigma$ uncertainties such that are consistent with our \cii{}-derived redshift at the 0.9 - 1.3 $\sigma$ level. It is instead possible that this UV emission emanates from background or foreground galaxies and spectroscopic follow-up, as will be available with JWST \citep[see][]{Stefanon2021}, is required to resolve this. We discuss the possible physical morphologies of REBELS-25 in more detail in Section~\ref{sec:nature}.

\subsection{Kinematics}

\begin{figure*}
	\centering
	\includegraphics[width=0.95\textwidth]{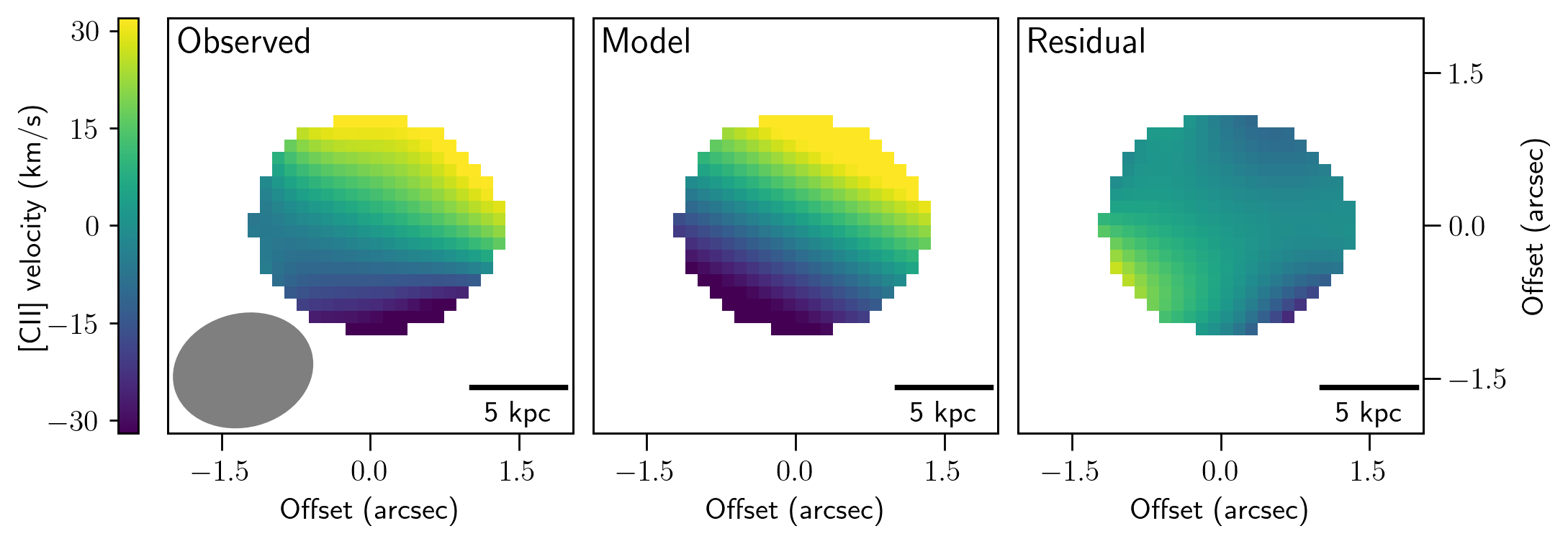}
	\caption{[Left panel] The \cii{} velocity field of REBELS-25  \citepschouws{}. The beam is shown as a grey ellipse in the bottom left corner of the Left panel and a scale bar showing 5 kpc is displayed in the bottom right corner of each panel. The velocity field exhibits a clear velocity gradient, which is indicative of disc rotation. We display the best-fitting thin disc model in the Middle panel and the residual between the model and the observed first moment map in the Right panel. However, we note that a similar fit with a merger model is statistically indistinguishable from this model.}
	\label{fig:mom1}
	
\end{figure*}

To visualise the kinematics of REBELS-25, we display a map of the \cii{} velocity field for REBELS-25 in Figure~\ref{fig:mom1}. For full details of the maps we refer the reader to  \citetschouws{}, where they are presented. In brief, the map is made by taking the first moment of the image, with a threshold that only includes individual pixels that are  above three times the noise level (i.e. $>3 \sigma$). The velocity field exhibits a velocity gradient across the galaxy. Such a velocity gradient could be indicative of disc rotation; however, the resolution of the data preclude us from distinguishing between this and other scenarios. 	Indeed,  \citetschouws{} fitted kinematic models of a thin, rotating disc shown in Figure~\ref{fig:mom1} and a separate model of an unresolved merger to REBELS-25 and found them to be statistically indistinguishable from one another. Thus, although we cannot definitively conclude on the source of the kinematic profile, both rotating disc and major merger models provide adequate fits to the data.

 Indeed, \citet{Rizzo2022} found that good resolution is essential to correctly model the kinematic state of simulated data, with at least three independent resolution elements across the major axis of the beam needed; this criterion is not met by our low-resolution data. It is plausible, especially given the fact that three separate rest-frame UV sources are detected \citep{Stefanon2019}, that the kinematic profile that we observe may be an artefact resulting from beam smearing of the kinematic profiles of separate sources \citep[see][]{Kohandel2020}. Higher resolution observations are necessary in order to distinguish between these scenarios. Finally, we characterise the velocity gradient in REBELS-25 by estimating the FWHM of the \cii{} line from a fit to the main \cii{} emission component with a single Gaussian, which results in a FWHM of $316 \pm 15$ \kms{}.

\subsubsection{Dynamical mass}
\label{sec:dynamical mass}

We proceed to estimate the dynamical mass, $M_{\rm dyn}$, of REBELS-25. We employ the method outlined in \citet{Kohandel2019} to calculate it as

\begin{equation}
M_{\rm dyn} = 2.35 \times 10^{9} \msol{}  \left(\frac{1}{\gamma^{2} \sin^{2}{i}}\right)  \left(\frac{\rm FWHM}{100~\kms{}}\right)^{2} \frac{R}{\rm kpc},
\end{equation}

\noindent
where the  \cii{} FWHM is measured in km s$^{-1}$, the radius of the galaxy, $R$, is measured in kpc, $i$ is the inclination of the galaxy and $\gamma$ is a parameter that depends on the physical nature of the source being considered, in particular its structure and kinematics. For REBELS-25, this translates to $M_{\rm dyn} = 8.5 \pm  1.9 \times 10^{10} \gamma^{-2} \sin^{-2}{i}  ~ \msol$, where we take $R$ to be the major axis of the main \cii{} component, as we have assumed that REBELS-25 is inclined. With our estimate of the inclination of the main \cii{} component of REBELS-25 that we obtained from Gaussian fitting of the component in the zeroth moment map (see Section~\ref{sec:morphology}), $i = \ciiinc{} \degr{} $, this translates to a dynamical mass $M_{\rm dyn}$ $ = 1.2  \pm  0.7 \times 10^{11} \gamma^{-2}  \msol{}   $

Due to the significant uncertainty in REBELS-25's morphology, the value of $ \gamma$ is similarly uncertain. \citet{Kohandel2019} give a range of values for $\gamma$ derived from Alth{\ae}a, a simulated high-redshift galaxy \citep{Pallottini2017}, with $\gamma$ = 1.78, 2.03, 1.52, when the galaxy is in a spiral disc, disturbed disc, and merger stage, respectively. Although simulated observations of Alth{\ae}a are made at a slightly lower redshift than REBELS-25 of $z=6$, it has a comparable stellar mass of $\sim 10^{10}~\msol$ and dynamical mass of $\sim 10^{10}~\msol$ \citep{Kohandel2019}. As REBELS-25 could reasonably be a disc in various kinematic states or a merger with the information available given the current resolution of our data, we do not select one value of $\gamma$ and we thus evaluate $M_{\rm dyn}$ for all three values and report $M_{\rm dyn}$ as a range $M_{\rm dyn} = 2-8 \times 10^{10} \msol$.

\subsection{ISM properties}
\label{sec:ism_properties}

We use our measurement of the \cii{} luminosity of REBELS-25 (see Section~\ref{sec:spectral_fitting}) to estimate the molecular gas mass, $M_{{\rm H}_{2}}$ in solar masses, of REBELS-25 from the empirical relation calibrated by \citet{Zanella2018},

\begin{equation}
	\label{eq:zanella18_mh2}
	M_{{\rm H}_{2}} = \alpha_{\cii{}}L_{\cii{}},
\end{equation}

\noindent
where $L_{\cii{}}$ is the \cii{} luminosity in solar luminosities and $\alpha_{\cii{}} = 31^{+31}_{-15} \msol \lsol^{-1}$ is an empirically calibrated conversion factor. From this relation, we calculate a molecular gas mass of $M_{{\rm H}_{2},\cii{}} = \mmolcii{} ~ \msol $. We note, however, that this relation was calibrated for galaxies at significantly lower redshift than REBELS-25 and as noted by \citet{Zanella2018} the conditions of the ISM at very high-redshift are most probably different to low-redshift conditions, which may impact the applicability of the relation to REBELS-25. However, an analysis of the molecular gas masses of a sample of galaxies from the ALPINE survey \citep{LeFevre2020, Bethermin2020, Faisst2020} between $4.4<z<5.9$ found that the \citet{Zanella2018} relation agreed with those derived from dynamical masses and dust continuum luminosities \citep{Dessauges-Zavadsky2020}. We have adopted $\alpha_{\cii{}} = 31^{+31}_{-15}$, following \citet{Zanella2018}, however, we note that there remains significant uncertainty as to the value of $\alpha_{\cii{}}$, with studies based on observations and simulations finding values between a few \citep{Rizzo2021}\footnote{We note that \citep{Rizzo2021} determine a conversion between \cii{} luminosity and the total gas mas, implying a conversion between \cii{} luminosity and molecular gas mass of at most a few. } and about a hundred \citep{Madden2020}. Simulations also suggest that low metallicity galaxies in particular may require a lower value of $\alpha_{\cii{}}$ \citep{Vizgan2022}.

\subsection{Star formation}
\label{sec:star_formation}

We calculate the total star formation rate for the galaxy using UV to trace unobscured star formation and IR to trace obscured star formation, as

\begin{equation}
	{\rm SFR}_{\rm tot} =  k_{\rm IR} L_{\rm IR} +  k_{\rm UV} L_{\nu, \rm UV},
\end{equation}

\noindent
where $k_{\rm IR} = 1.2\times 10^{-10} \msol \yr^{-1} \lsol^{-1} $ and $k_{\rm UV} = 7.1 \times 10^{-10} \msol \yr^{-1} ({\rm ergs~s^{-1} Hz^{-1}} )^{-1}  $. These values of $k_{\rm IR}$ and $k_{\rm UV}$ have been adopted for the REBELS sample \citep{Bouwens2022} and the rationale and calculation of these factors are further discussed by \citetstefanon{}, \citet{Inami2022} and \citet{Topping2022}.

\citet{Inami2022} determine an IR luminosity for REBELS-25 of $L_{\rm IR} = \lir{} $ \lsol{} (for further details see Section~\ref{sec:source_properties}). The UV luminosity of  REBELS-25 is calculated from the best-fitting SED-template obtained with {\sc EAzY} \citep{Brammer2008} to be $L_{\rm UV} = 9 \pm 2 \times 10^{10}~\lsol$ \citepstefanon{}. This translates to  ${\rm SFR}_{\rm tot} = \sfrtot{}~\msol~\yr^{-1}$, with ${\rm SFR}_{\rm UV} = \sfruv{} ~ \msol \yr^{-1}$ and ${\rm SFR}_{\rm IR} =  \sfrir{} ~\msol \yr^{-1}$. From this it can be seen that the obscured star formation, as traced by IR emission, dominates the star formation in REBELS-25.

We also make a second calculation of the SFR by using the \cii{} emission. \citet{DeLooze2014} calibrated a relationship between the \cii{} luminosity and star formation rate  based on a sample of galaxies, which they refer to as their high-redshift sample, $z = 0.5 - 6.6$

\begin{equation}
\log {\rm SFR}_{\cii{}} = -8.52 + 1.18  \log L_{\cii{}} + \log C_{\rm Kroupa} ,
\end{equation}

\noindent
where $C_{\rm Kroupa} = 1.08$ is a factor to convert from a \citet{KroupaWeidner2003} IMF to a \citet{Chabrier2003} IMF. From this, we calculate ${\rm SFR}_{\cii{}} = \sfrcii{} \msol {\rm yr}^{-1}$. 

 Both methods of calculating the star formation broadly indicate that REBELS-25 is undergoing an active phase of star formation in which a large amount of stellar mass is being added. However, the value of ${\rm SFR}_{\cii{}}$ is slightly ($\sim 1.25$ times)  greater than the value of ${\rm SFR}_{\rm tot}$. Although, when additionally taking account of the significant 0.4 dex uncertainty in the fitted relation this value is consistent with our measurement of the SFR from IR and UV. We note in general that calibrations for the SFR at such high redshift are uncertain due to the limited nature of the observations. Moreover, for \cii{} as a tracer of star formation specifically, \citet{Carniani2018} find that the relationship between SFR and $L_{\cii{}}$ holds in a sample of $z = 5.2 -7.2$ galaxies, but has significantly increased scatter in comparison to the relation exhibited in local galaxies. Similarly, \citet{Schaerer2020} find that the relationship between SFR and $L_{\cii{}}$ is valid in the galaxies of the ALPINE sample at redshifts $4.4 < z < 5.9$, but also with increased scatter compared to the local relation.  As discussed by \citet{DeLooze2014}, the scatter in the relation between SFR and \cii{} luminosity likely results, in large part, from the differing ISM conditions between different galaxies. Indeed, evidence from observations of local galaxies indicates that metallicity, which is uncertain for REBELS-25, particularly affects the fraction of \cii{} emission emanating from neutral and ionised media \citep{Croxall2017}. Therefore, although there is little evidence of evolution in the relation, the increased scatter at higher redshifts disfavours its usage compared to the UV + IR SFR. We thus adopt the SFR as calculated from the UV and IR emission (i.e. ${\rm SFR}_{\rm tot}$) for our further analysis.

We proceed to calculate a molecular gas depletion time, $ t_{\rm depl, {\rm H}_{2}}$, for REBELS-25, that is the time it would take for the inferred molecular gas reservoir of REBELS-25 to be depleted assuming a constant star formation rate and no other gain or loss of molecular gas mass. Under these assumptions, $  t_{\rm depl, {\rm H}_{2}} = M_{{\rm H}_{2},\cii{}} / {\rm SFR}_{\rm tot} = \tdepl{}$ Gyr. However, as we discuss in Section~\ref{sec:spectral_fitting} the additional secondary component that is observed in the \cii{} spectrum is evidence for either an outflow of gas, or a merger or inflow of gas. Under the assumption of no other effects, an outflow would lead to faster depletion of the gas reservoir whereas a merging galaxy or inflow of gas would increase REBELS-25's gas reservoir and lead to a longer depletion time.

\subsection{Properties of the secondary \cii{} emission component}
\label{sec:offset}

As discussed in Section~\ref{sec:spectral_fitting}, we observe an additional emission component in the REBELS-25 spectrum, slightly to the lower frequency-side of the main line component and fit a number of models to the spectrum of the galaxy. The two best-fitting models, shown in Fig~\ref{fig:cii_spectrum}, model the secondary  component as a merger or an outflow. In this section we present the physical properties that can be determined from our two best-fitting models.

\subsubsection{Merger model}
\label{sec:merger}

One natural origin for this secondary emission component is from another galaxy separate to REBELS-25 that may be merging with the main component.  We find that the centre of the secondary emission component is spatially coincident (to within 1.5 kpc) with the primary emission component from image-plane fitting (see also Figure~\ref{fig:mom0}) and is offset from the primary emission component in velocity space  by $\sim500~\kms{}$. This close association would suggest that the galaxy will merge into the primary emission component of REBELS-25. %

From this model, we measure a \cii{} luminosity of \lciimerger{}~\lsol{} for this merging galaxy. Using Equation~\ref{eq:zanella18_mh2}, this translates to a molecular gas mass of \mmolmerger{}~\msol{}, or about 18 per cent the mass of REBELS-25's molecular gas reservoir. If we assume that the total mass of the galaxies are proportional to their molecular gas reservoirs that we measure from \cii{}, this would classify this as a minor merger according to the commonly used definition of a minor merger as one where the mass ratio between the two galaxies is 1:3 or greater, however, we note that there can be significant differences between the ratios of total mass and the mass ratios of individual components of a galaxy \citep{Stewart2009}. Although only a minor merger, this would still serve to increase the amount of molecular gas in REBELS-25's reservoir and thus the galaxy's potential future stellar mass.

\subsubsection{Outflow model}
\label{sec:outflow}

Another plausible origin for the secondary emission component is from an outflow. We follow the method outlined by \citet{Herrera-Camus2021} to estimate the properties of our best-fitting outflow model. The atomic hydrogen outflow rate $\dot{M}_{\rm H{\sc I}, out}$, can be estimated as

\begin{equation}
\dot{M}_{\rm H{\sc I}, out} = M_{\rm H{\sc I}, out} \left( \frac{ v_{\rm out}}{R_{\rm out}}  \right),
\end{equation}

\noindent
where $M_{\rm H{\sc I}, out}$ is the atomic hydrogen mass of the outflow, $v_{\rm out}$ is the velocity of the outflow and $R_{\rm out}$ is the size of the outflow.

Following \citet{Herrera-Camus2021}, we first calculate an estimate of the projected outflow velocity, $v_{\rm out, proj}$, as

\begin{equation}
	v_{\rm out, proj} \sim \left| \Delta v \right| + \frac{ FW_{10\%}}{2} ,
\end{equation}

\noindent
where $\Delta v$ is the difference in velocity between the outflow and the main component and $FW_{10\%}$ is the full width at tenth maximum of the outflow component. The best-fitting outflow component of our model has a $FW_{10\%} = 1600 \pm 400 \kms{}$  and is separated from the main \cii{} emission component by $\left| \Delta v \right| = 100 \pm 80 $ \kms{}, which translates to a projected outflow velocity of $900 \pm 200$ \kms{}.

Under the assumption of an outflow perpendicular to the assumed disc of REBELS-25, we can use our estimate of the inclination of the main \cii{} component, $i =  \ciiinc{} \degr{}$ to further estimate the deprojected  outflow velocity as $v_{\rm out, deproj} =  v_{\rm out} / \cos{i} = 1600 \pm 300$ \kms{}.

\citet{Herrera-Camus2021} presented a relationship between atomic hydrogen mass in the outflow, $M_{\rm H{\sc I}, out}$ and the \cii{} luminosity under the assumption that the collisional excitation of \cplus{} is primarily due to hydrogen atoms:

\begin{equation}
	M_{\rm H{\sc I}, out} = \kappa_{\scriptsize \cii{}} L_{\scriptsize \cii{}} ,
\end{equation}

\noindent
where $\kappa_{\cii{}}$ is a conversion factor dependant on the temperature, metallicity and number density of atomic hydrogen gas. We calculate a lower limit to the atomic gas mass under the ``maximal excitation conditions'' scenario presented by \citet{Herrera-Camus2021},  with $\kappa_{\cii{}} = 1.5~\msol / \lsol$. This value of $\kappa_{\cii{}}$ is calculated by assuming a Helium fraction of 36\%, solar metallicity, a temperature of $10^4$ K and number density of $ 10^{4}~{\rm cm}^{-3} $ for the gas. We determine a \cii{} luminosity of $7 \pm 3 $  $\times 10^8$ \lsol{} for the outflow component of our best-fitting model, which translates to an estimated lower limit of  $M_{\rm H{\sc I}, out}$ $ \gtrsim 1.1 \times 10^9$ \msol{}.

We have chosen to use the ``maximal excitation conditions'' scenario presented by \citet{Herrera-Camus2021} in order to provide a conservative estimated lower limit of the outflow mass. However, we note that the true physical conditions of the gas may well be different to those that we have considered. In particular, although measurements of the metallicity of high redshift galaxies are limited, subsolar metallicities are considered to be most likely  \citep[see reviews by][]{Stark2016,DayalFerrara2018}. With the other conditions kept the same, a subsolar metallicity would lead to an increase of$\kappa_{\cii{}}$ to $4.7~\msol / \lsol$ in the case of half solar metallicity and $41.7~\msol / \lsol$ in the case of metallicity one tenth of solar metallicity \citep[for the details of the calculation of these values of $\kappa_{\cii{}}$ see][]{Herrera-Camus2021}. This would lead to a corresponding increase in our estimate of the lower limit of the outflow mass by a factor of $\sim3$ for half solar metallicity and a factor of $\sim30$ in the case of one tenth solar metallicity. In order to provide a better estimate of the outflow mass, multi-line observations of REBELS-25 are required to obtain an accurate estimate of the metallicity.

Last, we use the average of the major and minor axes of our image plane fit to the outflow component (see Section~\ref{sec:morphology})  to estimate $R_{\rm out} \approx 9$ kpc.  We then combine our estimates of $M_{\rm H{\sc I}, out}$, $v_{\rm out}$ and $R_{\rm out}$ to derive an estimated lower limit of the  projected atomic mass outflow rate of $\dot{M}_{\rm out, proj} \gtrsim 120$ \msol{} $\yr^{-1}$, and an estimated lower limit of the deprojected atomic mass outflow rate  $\dot{M}_{\rm out, deproj} \gtrsim \moutdproj{}$ \msol{} $\yr^{-1}$.  From these outflow rates we estimate a lower limit of the atomic mass loading factor, $ \lambda =  \dot{M}_{\rm out} / {\rm SFR} $. We estimate a projected mass loading factor of  $ \lambda_{\rm proj}  \gtrsim 0.6$ and a deprojected mass loading factor of  $ \lambda_{\rm deproj}  \gtrsim 1.0$. In other words we estimate a project lower limit that is roughly half the SFR of REBELS-25 and a deprojected lower limit that is roughly equivalent to this SFR.

\section{Discussion}
\label{sec:discussion}

\subsection{The Nature of REBELS-25}
\label{sec:nature}

We discuss here the nature of REBELS-25 in the context of the observational results that we have presented in Section~\ref{sec:methods}  and results from the literature.

\subsubsection{ULIRG}
\label{sec:ulirg}

REBELS-25 is notable for its significant IR emission ($L_{\rm IR} = \lir{}$ \lsol{}), making it the only ULIRG in the REBELS sample (see Figure~\ref{fig:rebels_sample}). If the ISM conditions that we have adopted in order to determine the IR luminosity deviate from the true ISM conditions of REBELS-25, this will in turn impact our estimate of the IR luminosity. In particular, we have adopted the median dust temperature (46K) determined from a sample of the REBELS galaxies (see Section~\ref{sec:source_properties} for further details). Direct determinations of the dust temperature by SED fitting in galaxies at a redshift comparable to REBELS-25 are limited by the lack of sources with observations at higher frequency than the dust emission peak. However, the temperature that we adopt is comparable to the temperature of $42^{+13}_{-7}$K measured for a  less IR luminous ($ \sim 2 \times 10^{11} \lsol{}$) galaxy  at $z = 7.13$ that has such observations \citep{Bakx2021}.

There have now been two works that attempt to determine the dust temperature and IR luminosity of REBELS-25 in the absence of a well-sampled dust SED. Using their method combining a single dust continuum measurement with a \cii{} line luminosity, \citet{Sommovigo2022} determine an even higher dust temperature  ($55^{+15}_{-14}$K, corresponding to an IR luminosity of $\log{L_{\rm IR}}   = 12.45^{+0.43}_{-0.45} \lsol$) than that which we have  adopted. Meanwhile, using an additional ALMA Band 8 observation of REBELS-25, \citet{Algera2023} fit an SED to the two observed  continuum fluxes.  In their fiducial optically thin model with a dust emissivity index $\beta = 2.0$, they determine a lower dust temperature of $34^{+6}_{-6}$K and a corresponding IR luminosity of  $\log{L_{\rm IR}}   = 11.85^{+0.18}_{-0.10}\lsol$. However, when modifying their fiducial model by instead adopting $\beta = 1.5$ or fitting for $\beta$, they determine temperatures of $43^{+11}_{-8}$K and $42^{+24}_{-11}$K, corresponding to  IR luminosities of $\log{L_{\rm IR}}  = 12.02^{+0.28}_{-0.20}\lsol$ and $\log{L_{\rm IR}}  = 12.00^{+0.46}_{-0.22}\lsol$, respectively. They find even higher values for the dust temperature and IR luminosity in the optically thick scenarios compared to the corresponding optically thin scenarios.

Given its probable status as a ULIRG, the question arises as to whether REBELS-25 has different dust properties to the rest of the REBELS sample and what impact this may have on the IR luminosity. \citet{Sommovigo2022a} find that the ULIRGs in the sample of ALPINE galaxies that they analyse have higher dust temperatures than the other galaxies. However, the other conditions of the dust, in particular the dust emissivity index, play an important role, as seen above. Observations of ULIRGs in the Local Universe \citep{Clements2018} and $z~\sim~1-3$ submm galaxies \citep{DaCunha2021} indicates that there is a general correlation between higher dust temperature and higher IR luminosity for high IR luminosity galaxies (including ULIRGs). Furthermore, observations of nearby galaxies \citep{Lamperti2019} and \mbox{$z~\sim~1-3$} submm galaxies \citep{DaCunha2021} show a correlation between higher dust temperature and lower dust emissivity index. If these trends are also present in the Epoch of Reionisation then this would further support a higher IR luminosity for REBELS-25, in the case that we have adopted a lower dust temperature than the true value. If we have underestimated the IR luminosity of REBELS-25, we would have consequently underestimated the obscured (and thus total) SFR. To conclusively determine the IR luminosity of REBELS-25 and thus its SFR, high frequency observations of REBELS-25 that well-sample the dust SED are required.

Recent work by \citet{Zavala2021} to constrain the IR luminosity function, suggests that ULIRGs should make the dominant contribution to the total  obscured star formation rate  at the redshift covered by the  REBELS sample. However, the other (non-ULIRG) galaxies in the REBELS sample account for the majority of the obscured star formation in the REBELS sample  \citep{Inami2022}. It is possible, however, that this discrepancy  is explained by the design of the REBELS survey. The REBELS survey selected galaxies on the basis of UV observations \citep{Bouwens2022} and thus may be prone to disproportionately select dust-poorer galaxies for observation, which are less likely to have significant IR luminosities in comparison to the population of galaxies as a whole.  Indeed this scenario is also a possible explanation for the tension between the dust masses of the REBELS sample and the UV luminosity function  \citep[for further discussion see][]{Dayal2022}. This scenario is supported by the finding of \citet{Sommovigo2022} who present an analytical model of the evolution of dust temperature with redshift and find that the dust temperature of the REBELS galaxies is typical of quite UV-transparent galaxies in the context of this model.

Furthermore, the IR luminosities of the REBELS galaxies presented by  \citet{Inami2022} and used in this paper are calculated by converting a monochromatic luminosity to a total IR luminosity.	There is of course uncertainty inherent in this conversion, including from the uncertainty in the dust temperature \citep[see discussion in][]{Sommovigo2022}. It is possible that a more-thorough determination of their IR luminosities from observations that sample the IR luminosity distribution at multiple wavelengths would result in a greater fraction of the REBELS galaxies being classified as ULIRGs. The evidence from other approaches to constrain the IR luminosity is mixed, however. \citet{Sommovigo2022} used a model to predict the IR luminosities of a subsample of thirteen of the REBELS galaxies using both their monochromatic continuum luminosities and \cii{} line luminosities; they predict REBELS-25 to be the only ULIRG in the subsample. \citet{Ferrara2022} used UV continuum flux, the UV spectral slope and the monochromatic continuum luminosity of the REBELS galaxies in combination with radiative transfer models using different simple geometries to derive IR luminosities of the sample; for a Milky Way extinction curve, they find two other galaxies in the REBELS sample that have IR luminosities above $10^{12}$~\lsol{}. However, for an SMC extinction curve they find IR luminosities below  $10^{12}$~\lsol{} for both galaxies. We note that \citet{Ferrara2022} could not determine an IR luminosity for REBELS-25, due to its very high dust-continuum flux, which is consistent with it being very  IR luminous in comparison to the REBELS sample.

\subsubsection{Morphology}

Our observations indicate a morphologically complex nature for REBELS-25. The spectrum of REBELS-25 includes a faint \cii{} emission component, spectrally-separated from the main \cii{} emission component (See Figure~\ref{fig:cii_spectrum}), which we have interpreted as being potential evidence of a minor merger or an outflow (see Section~\ref{sec:spectral_fitting}). Although there are still limited constraints on the frequency of mergers at high redshift, they appear somewhat common, but less frequent than at intermediate redshifts. A recent analysis also suggests that the fraction of minor mergers decreases towards higher redshift after peaking at intermediate redshift, with the fraction of minor mergers being 8-13 per cent at  $ z \geq  3$ \citep{Ventou2019}, consistent with prior analyses that indicate that the fraction of major mergers decrease at higher redshift \citep{Xu2012,Ventou2017,Ventou2019}. \citet{LeFevre2020} found that 40 per cent of galaxies in the ALPINE sample of galaxies at $4 < z < 6$ were mergers with a subsequent analysis combining the ALPINE survey with lower redshift samples by \citet{Romano2021} indicating that the peak of the major merger rate is at $z\sim 3$ and slowly declines towards higher redshifts, with a major merger frequency of 34 per cent at $z\sim 5.5$. 
Star-formation-driven outflows have also been detected at high redshift \citep[e.g.][]{Herrera-Camus2021}, though statistics on their frequency remain somewhat scant. Theoretical results suggest, however, that \cii{} haloes can be explained by prior outflows \citep{Pizzati2020}, which would provide evidence that they are common at this redshift. We explore the observational context for and implications of an outflow further in Section~\ref{sec:outflow_implications}, but note that both scenarios are plausible.

Prior analysis of rest-frame UV imaging identified multiple UV components at the location of REBELS-25 \citep{Stefanon2019} and a prior study found dust continuum emission offset from these UV components \citep{Schouws2022a}. We observe an offset between this UV emission and both the centres of the  \cii{} and dust continuum emission in REBELS-25 (see Figure~\ref{fig:mom0}). In addition, \citet{Ferrara2022} finds that REBELS-25 has a much larger ratio of dust continuum flux to UV continuum flux than would be expected from its UV spectral slope, which is indicative of UV and IR emission emanating from spatially decoupled regions. Offsets between the UV and dust emission in galaxies have been commonly observed in lower-redshift SMGs and may indicate  a physical offset between the dust and the sites of star formation, for example due to the morphological disturbance of a merger or may instead suggest that the UV clumps are embedded in a larger dusty galaxy and are visible due to reduced attenuation from dust at their location \citep{Hodge2015,Hodge2016,Chen2017,CalistroRivera2018,Hodge2019,Rujopakarn2019,Cochrane2021}. At high-z, offsets between \cii{} and UV emission have also been observed. \citet{Maiolino2015}  found such an offset in three $z\sim 7$ galaxies and they attributed the lack of \cii{} emission in the vicinity of the UV emission to molecular clouds being dispersed by stellar feedback on short timescales. Although the observed offset between the centre of the \cii{} emission and the UV emission components could hint at a similar scenario in REBELS-25, the large beam of our current \cii{} data means that it is not possible to determine whether or not there is \cii{} emission coincident with the UV emission. \citet{Fujimoto2020} measured offsets between the UV and both the \cii{} and dust continuum for the galaxies of the ALPINE sample at $4 < z < 6$. For those galaxies not identified as mergers, they were unable to determine whether the observed offsets were physical or a result only of astrometric and positional uncertainty. More recently, two studies have examined the frequency of offsets between dust and UV emission in the EoR. \citet{Schouws2022a} found that the dust and UV were co-spatial in the majority of the six galaxies at $z \sim 7 - 8$ that they detected in dust, with the exception of REBELS-25, for which they observed an offset. \citet{Bowler2022} found that three out of a sample of six bright galaxies at $z\sim 7$ with high resolution ALMA imaging have a significant dust detection that is offset from the galaxy's UV emission. Offsets between the dust continuum emission and UV are explored for the REBELS sample in \citet{Inami2022}, who find additional galaxies within the sample that exhibit offsets. Thus the picture at high-z remains somewhat unclear, but initial results suggest that offsets between dust and other components are somewhat common in bright galaxies at high redshift.

\subsubsection{Kinematics}

We observe a  significant velocity gradient in the main \cii{} emission component of REBELS-25 (see Figure~\ref{fig:mom1}), which may be indicative of a rotating disc. However, with the large size of the beam of our observations, the observed velocity gradient could also be produced by beam smearing of merging galaxies. As discussed above, mergers are not uncommon at high redshift. It is unclear, however, how common rotating discs are at such high redshifts.  The conventional model of disc formation in galaxies, has discs forming through the process of cooling and dissipation \citep[see e.g][]{Eggen1962,ReesOstriker1977,WhiteRees1978,FallEfstathiou1980}, which results in the formation of discs at relatively late times, roughly around the time of cosmic noon. However, signatures of rotating discs have been observed out to higher and higher redshifts $z> 4-5$  \citep[see e.g.][]{Smit2018,Pavesi2019,Bakx2020,Rizzo2020,Neeleman2020,Jones2021,Lelli2021}. Indeed, in a study of the kinematics of high-z ($z\gtrsim 6$) quasar host galaxies, \citet{Neeleman2021} found ten out of the 27 studied galaxies have a rotational profile consistent with that of a rotating gas disc. In addition, from the perspective of theory, modern, hydrodynamical simulations also are able to produce a significant fraction of disc galaxies at redshifts well-above cosmic noon \citep[e.g.][]{Pallottini2017, Pallottini2019, Pillepich2019, Leung2020,Kohandel2020}.

Overall, the observational evidence indicates that REBELS-25 is a morphologically complex galaxy, potentially comprising some combination of a rotating disc, merging galaxies and outflowing gas. In addition, REBELS-25 is observed to have significant stellar, gas and dust mass components built up at a high redshift. However, further multi-wavelength and higher resolution observations are required to resolve the uncertainties in the morphology and composition of the galaxy. %

\subsection{Main sequence comparison}
\label{sec:ms_comparison}

We can compare the star formation properties of REBELS-25 to the star formation main sequence of \citet{Speagle2014} and the relation for $ t_{\rm depl, {\rm H}_{2}} $ of \citet{Liu2019}. We note, however, that as both \citet{Liu2019} and \citet{Speagle2014} based their studies on data out to redshift $z \sim 6$, our application of these relations represents an extrapolation of these data under the assumption that the trends observed out to $z \sim 6$ will continue to hold at higher redshift, which may not be valid.

The best-fitting main-sequence of \citet{Speagle2014} is

\begin{equation}
\log{{\rm SFR}_{\rm MS}} = (0.84 - 0.026  t_{\rm age}) \log{M_{*}} - (6.51   -  0.11  t_{\rm age}),
\end{equation}
\noindent
where $t_{\rm age}$ is the age of the universe in Gyr, which at $z \sim \zciitwodp{}$ is 0.71 Gyr with our adopted cosmology. For the stellar mass of REBELS-25 (see Section~\ref{sec:source_properties}) this translates to a prediction of ${\rm SFR}_{\rm MS} = \sfrms{}~$ \msol yr$^{-1}$. The total SFR that we measure for REBELS-25 of $\sfrtot{}~\msol \yr^{-1}$ is approximately four times this main-sequence value of \citet{Speagle2014}. If we instead take the larger non-parametric stellar mass estimate, this would translate to a prediction of \sfrmsnonpar{}~\msol yr$^{-1}$, placing REBELS-25's SFR at about twice the main-sequence value.  \citet{Topping2022} use the REBELS sample to determine a high-redshift main-sequence by fixing the slope of the main-sequence to that presented by \citet{Speagle2014} and selecting the best-fitting intercept. The main sequence presented by \citet{Topping2022} was determined using a non-parametric method to determine stellar mass. If we adopt the same approach as \citet{Topping2022}, but instead use stellar masses for the REBELS sample that have been determined by assuming a constant star formation history, as adopted in this paper, we determine a main sequence of the form

\begin{equation}
	\label{eq:topping_ms}
	\log{{\rm SFR}_{\rm T22}} =  0.82 \log{M_{*}} - 5.93.
\end{equation}

\noindent
Using this prescription would result in a higher main sequence SFR of ${\rm SFR}_{\rm T22}$ $\sim 151$ \msol   yr$^{-1}$. REBELS-25's SFR is comparable to (1.3 times) this predicted value. 

\citet{Khusanova2021} presented a main sequence derived from the ALPINE survey of galaxies. We consider the main sequence presented for galaxies at $z~\sim5.5$, which is the higher redshift of the two main sequences presented by \citet{Khusanova2021}. This main sequence has the form

\begin{equation}
	\label{eq:khusanova_ms}
	\log{{\rm SFR}_{\rm K21}} =  0.66 \log{M_{*}} - 5.1
\end{equation}

\noindent
(\citealt{Khusanova2021}; Khusanova, private communication).\footnote{The slope of the main sequence was presented in \citep{Khusanova2021} and the intercept was obtained via private communication.} Using this prescription would result in a lower main sequence star formation rate of ${\rm SFR}_{\rm K21} \sim 27$ \msol yr$^{-1}$ with our adopted stellar mass. In comparison REBELS-25's SFR is $\sim 7$ times this value. If we instead use the non-parametric stellar mass of REBELS-25 this would result in a main sequence SFR of $ \sim 48$ \msol yr$^{-1}$, placing REBELS-25  $\sim4$ times above this main sequence value.

Although there is no consensus regarding the definition of a starburst galaxy, a simple definition is a threshold based on a multiple of the main-sequence-predicted star formation rate. For example, if we take a threshold of three times in excess of that predicted by the main-sequence \citep[see e.g.][]{Elbaz2018}, this places REBELS-25 on the threshold for a starbursting galaxy if we adopt the \citet{Speagle2014} main sequence with our adopted stellar mass. However, given the range of star-forming main sequences and starburst definitions that could be adopted, such a classification must be seen as very tentative. Indeed, if we were to take the main sequence of Equation~\ref{eq:topping_ms}, or the non-parametric stellar mass presented by \citet{Topping2022}, REBELS-25 would be classified as a main sequence galaxy. In further contrast, if we take the main sequence prescription of \citet{Khusanova2021}, this would place REBELS-25 clearly in the starburst regime for both our adopted stellar mass and the non-parametric stellar mass.

\citet{Sommovigo2022} determined a dust mass of $4^{+4}_{-1} \times 10^{7}~\msol{}$  for REBELS-25, with a higher dust temperature and infrared luminosity than that adopted in this paper. This would correspond to a dust-to-gas ratio of 0.0007.  If we use the same dust model as \citet{Sommovigo2022}, but compute the dust mass with $T_{\rm dust} = 46$~K and $L_{\rm IR} = \lir{}~\msol$, as used throughout this paper, we would determine a lower dust mass of $\sim 2 \times 10^{7}~\msol{}$, which would correspond to a dust-to-gas ratio of 0.0004. The dust-to-gas ratio is strongly metallicity dependent in nearby galaxies \citep[see e.g.][]{Leroy2011,DeVis2019}, thus follow up observations to obtain a reliable measurement of the metallicity in REBELS-25 and other galaxies in the EoR would enable REBELS-25 to be placed in context.

Compared to local star-forming galaxies, which have a molecular gas depletion time of about $\sim 2$ Gyr \citep{Bigiel2008, Bigiel2011}, REBELS-25 has a much shorter molecular gas depletion time of $ t_{\rm depl, {\rm H}_{2}} =  \tdepl{} $ Gyr. Such comparatively short depletion times are common in high-z galaxies \citep[see e.g. review by][]{HodgedaCunha2020}. We calculate a predicted main-sequence depletion time of 0.53 Gyr\footnote{Calculating the main-sequence depletion time with the non-parametric stellar mass, results in a value of 0.55 Gyr, which is only very slightly different from the value calculated with our adopted stellar mass.} based on the prescription of \citet{Liu2019}. Our inferred value of $ t_{\rm depl, {\rm H}_{2}} $ is about one half of this predicted value, which is slightly longer than would be predicted from the enhanced SFR of REBELS-25, suggesting REBELS-25 also has a more massive gas reservoir than would be predicted by the main sequence.

\subsection{Implications of a potential outflow}
\label{sec:outflow_implications}

Outflows have been inferred from \cii{} in AGN host galaxies at a similar redshift as REBELS-25 \citep{Maiolino2012,Cicone2015,Izumi2021} and in star-forming galaxies at lower $4 <z < 6 $ redshift \citep{Gallerani2018,Ginolfi2020,Herrera-Camus2021}. Moreover, observations with other tracers indicate they are a common feature at high redshift, with 73 per cent of a sample of eleven lensed dusty star forming galaxies observed having a detected outflow \citep{Spilker2020a}. Outflows can transport material away from their host galaxies and thus can contribute to quenching star formation if the material escapes the galaxy. Modelling of observations indicates that outflows are a necessary component to explain observed quenching  \citep{Trussler2020}. They may also be the origin of \cii{} haloes around galaxies \citep{Pizzati2020}. They thus plays an important role in shaping the evolution of galaxies. If REBELS-25 has a sustained outflow over a long period, this would serve to reduce the gas available to form stars and thus the potential for REBELS-25 to accumulate stellar mass.

Observations in the Local Universe find that star formation-driven outflows typically have mass loading factors  $\lesssim 4$, whereas AGN-driven outflows can have much higher mass loading factors \citep[see e.g.][]{Cicone2014,Fluetsch2019}. Although observations of star-formation driven outflows at high-z are much more limited in comparison they typically find similarly low mass loading factors \citep[see e.g.][]{Ginolfi2020a,Spilker2020a,Herrera-Camus2021}.

\begin{figure}
	\centering
	
	\includegraphics[width=\columnwidth]{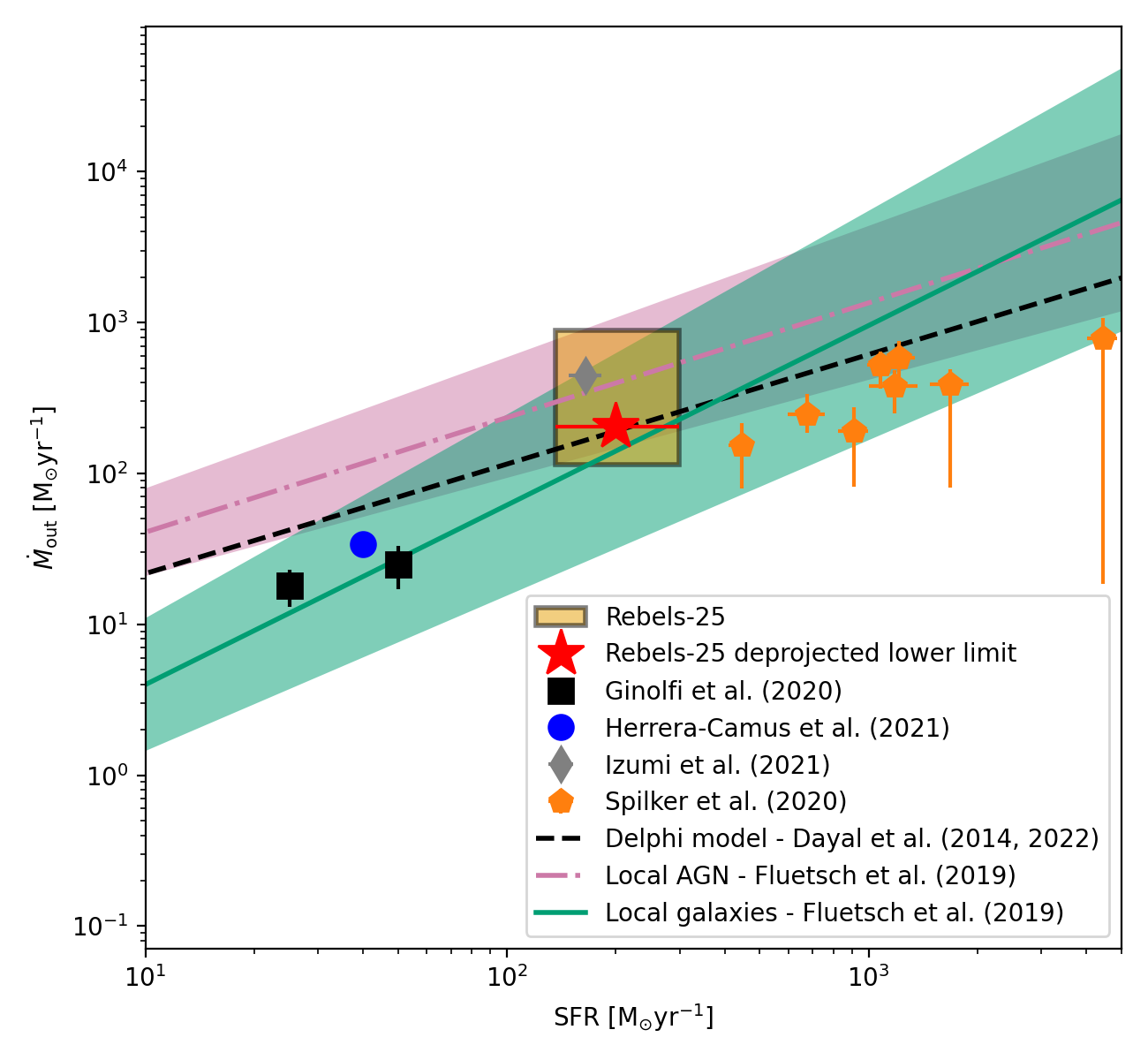}
	\caption{The outflow mass flow rate $\dot{M}_{\rm out}$ against the SFR for REBELS-25 and other sources from the literature.  We indicate the deprojected lower limit of $\dot{M}_{\rm out}$ for REBELS-25 with a red star, and we indicate the range of uncertainty as an orange rectangle with a black outline. The outflow properties of a $z\sim 5.5$ galaxy \citep{Herrera-Camus2021} are shown as a blue circle without errors, as it is reported as an estimate. Outflow properties from a sample of $z>4$ dusty star forming galaxies \citep{Spilker2020} are shown as orange pentagons with errors. The outflow properties measured from a stacking analysis of galaxies at $4 < z<6$  \citep{Ginolfi2020a} are show as black squares with errors. An outflow from a $z\sim 7$ AGN  \citep{Izumi2021} is shown as a grey diamond with errors on the SFR, but none on $\dot{M}_{\rm out}$, as this is reported as an estimate. The prediction of the \textsc{delphi} semi-analytic model \citep{Dayal2014, Dayal2022} is shown as a black dashed line. Scaling relations from \citet{Fluetsch2019}, which we have partially extrapolated beyond the highest SFR values probed in the \citet{Fluetsch2019} sample,  are shown for the outflow rate of local AGN-driven outflows (pink dot-dashed line and pink shaded area) and for the outflow rate of local star-forming galaxy driven outflows (solid green line and green-shaded area). Overall, the majority of the range of $\dot{M}_{\rm out}$ we estimate is consistent with local star-formation driven outflows, but  the upper range of our $\dot{M}_{\rm out}$ estimate  is more comparable to local AGN-driven outflows. }
	
	\label{fig:mass_loading }
\end{figure}

In Section~\ref{sec:outflow}, we estimated a lower limit on the mass loading factor of 0.6 (projected) and 1.0 (deprojected). A mass outflow rate slightly below the SFR (the projected lower limit) or roughly equivalent to it (the deprojected lower limit) would be consistent with a star-formation driven outflow. However, a large uncertainty in the determination of the mass loading factor is the size of the outflow. For these estimates, we adopted a size of 9  kpc based on our image plane fitting. However, it may be that the true size of the outflow is smaller. If we instead adopted a value of $R_{\rm out} = 6 $ kpc, as was used by \citet{Ginolfi2020a}, this would result in a projected mass loading factor of 0.8 and a deprojected mass loading factor of 1.5. In other words an outflow rate  that is comparable to or slightly above the star formation rate. Instead adopting $R_{\rm out} = 2 $ kpc, as was determined for an outflow by  \citet{Herrera-Camus2021}  with higher resolution observations, would result in a projected mass loading factor of 2.5 and a deprojected mass loading factor of 4.4. In other words, an outflow rate that is significantly in excess of the star formation rate and pushing towards values more comparable to mass loading factors observed in AGN-driven outflows \citep[see e.g.][]{Cicone2014,Fluetsch2019}. Evidence from local galaxies indicates that outflows typically have a total mass outflow rate about three times that of the atomic mass component for star formation driven outflows and two times as much for AGN-driven outflows \citep{Fluetsch2019}. If the relationship observed by \citet{Fluetsch2019} holds at high redshift, our largest estimates of the total outflow rate would be outside the $1 \sigma$ range of observations from local star-formation driven outflows, for the lower end of our star formation estimate for REBELS-25. However, the majority of our estimate range is consistent with the properties of local star formation-driven outflows. We display the range of mass outflow rates that we estimate with the above considerations, in comparison to a number of literature sources in Figure~\ref{fig:mass_loading }. Further work is needed to characterise star-formation driven outflows at high redshift to better understand their properties in comparison to those in the Local Universe.

We also compare our results to the predictions of the \textsc{delphi} semi-analytic model for galaxy formation \citep{Dayal2014, Dayal2022}. In brief,  \textsc{delphi} uses a binary merger tree approach to jointly track the build-up of dark matter halos and their baryonic components (gas, stellar, metal and dust mass). The model follows the assembly histories of $z \sim 4.5$ galaxies with halo masses ${\rm log}(M_h/ \msol)=8-14$ up to $z \sim 40$. The model contains only two mass- and redshift-independent free parameters, which are tuned to simultaneously reproduce the observed stellar mass function and the UV luminosity function at $z \sim 5-12$.  These two parameters are the maximum (instantaneous) star formation efficiency  (8 per cent) and the fraction of the type II supernova explosion energy (7.5 per cent) that is available to drive an outflow. For the full details of the physics underpinning the model we refer the reader to \citet{Dayal2014} and \citet{Dayal2022}.

For galaxies with ${\rm SFR}  > 1 ~\msol ~ {\rm yr^{-1}}$  the  \textsc{delphi} model yields an outflow rate ($\dot M_{\rm out}$) that scales with SFR as

\begin{equation}
	\log \dot M_{\rm out} = 0.068~\log\left( {\rm SFR}^2 \right) + 0.59~\log \left({\rm SFR}\right) + 0.61.
\end{equation}

\noindent
For REBELS-25's SFR of $\sfrtot{}~\msol ~ {\rm yr^{-1}}$ , this model would yield an outflow rate of $190^{+65}_{-45}~\msol ~ {\rm yr^{-1}}$. This prediction lies within the region of our observational estimate, which we display in Figure~\ref{fig:mass_loading } and is in good agreement with our observational estimate of the lower limit of the outflow rate of $\dot{M}_{\rm out, deproj} \gtrsim \moutdproj{}$ \msol{} $\yr^{-1}$.

\subsection{The fate of REBELS-25}
\label{sec:fate}

 Given its large stellar mass ($M_{*} =  \mstar{}~\msol{}$), large molecular reservoir ($\mmolcii{} ~ \msol $), high SFR ($\sfrtot{}  ~ \msol~\yr^{-1}$) at a redshift of $z = \zciitwodp{}$, one can ask whether REBELS-25 could evolve into a galaxy similar to massive high-redshift quiescent galaxies, such as those observed by recent studies \citep{Glazebrook2017,Schreiber2018a,Merlin2019,Carnall2020,Forrest2020,Forrest2020a,Saracco2020,Tanaka2019,Valentino2020,Kubo2021}  with stellar masses $\log M_{*} \gtrsim 10.5$ and significantly sub-main-sequence SFRs.

 Indeed, if we assume that REBELS-25 continues to form stars at its current rate with no other gain or loss of molecular gas and that it converts 100 per cent of its gas reservoir into stellar mass, REBELS-25 would deplete its molecular gas by $z \sim 5.8$ and have a total stellar mass of {$\log \left( M_{*} / \msol \right) = 10.8$}. This simple analysis suggests that REBELS-25 has the required molecular gas to obtain a stellar mass consistent with those of the recently observed high-redshift quiescent galaxies by $z\sim4$. However, the assumption of a constant SFR is unrealistic due to the very high star formation efficiency this would require at later times (without replenishment of the gas reservoir). Furthermore, observational evidence suggests that outflows are common in high-z dusty galaxies \citep[see e.g.][]{Spilker2020a} and we observe the potential signature of an outflow in REBELS-25 (see Section~\ref{sec:outflow}).

 We note that we have conservatively adopted a stellar mass of $M_{*} =  \mstar{} \msol{}$ for this analysis, as throughout the paper. However, if we instead adopted the higher non-parametric stellar mass of \nonparmass{}~\msol{} for this analysis, this would serve to slightly increase the model's predicted stellar mass for REBELS-25, but would not qualitatively alter our result.

 In this section, we investigate the feasibility of this scenario by considering a simple, conservative evolutionary model in which REBELS-25 evolves with only its current gas reservoir.

\begin{figure*}
	\centering
	\includegraphics[width=0.95\textwidth]{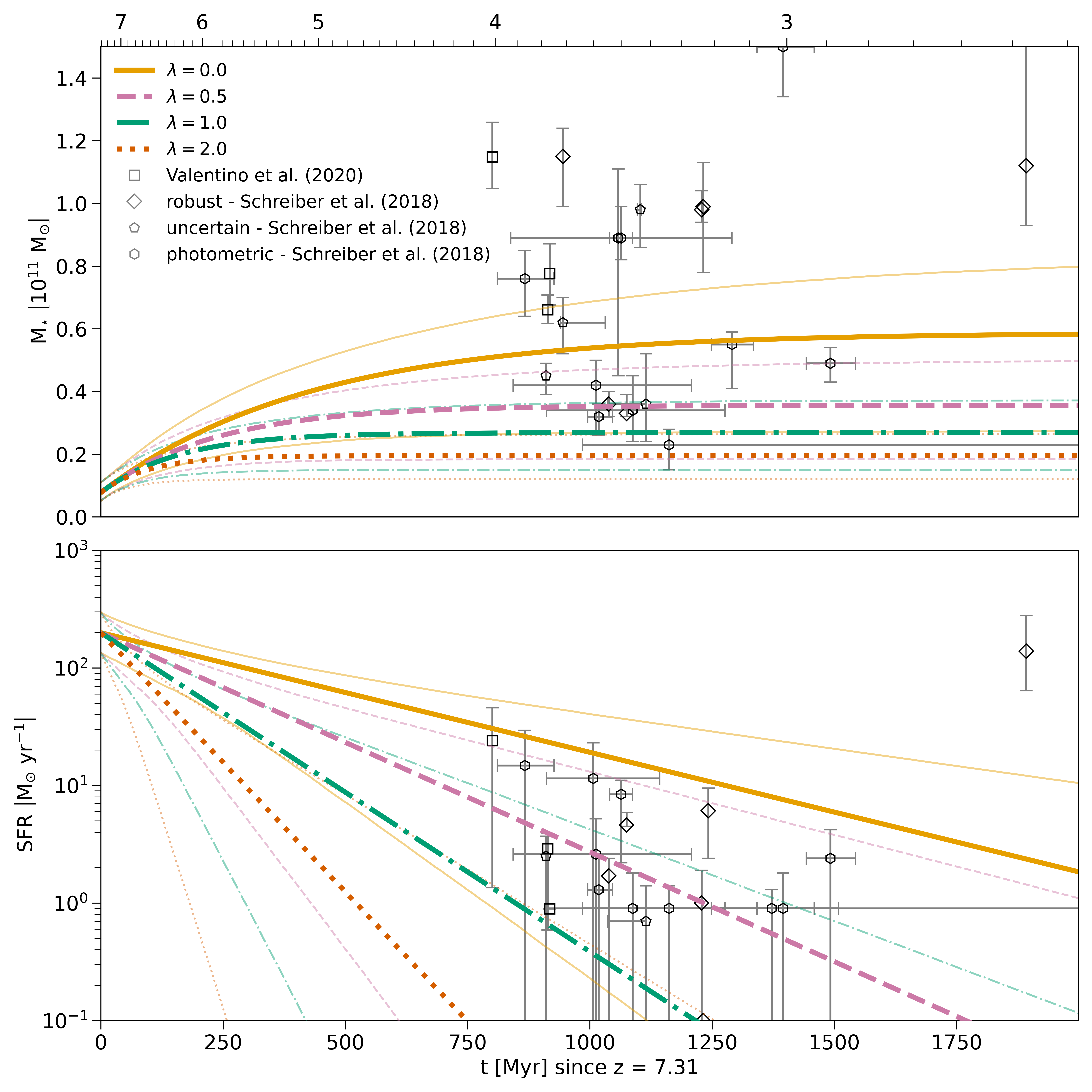}
	\caption{The predicted evolution of the stellar mass [Top panel] and SFR [Bottom panel] for REBELS-25 derived from a simple conservative model under the assumption of continuous star formation with constant star formation efficiency and no inflow of gas. We consider four different scenarios for a continuous gas outflow with mass loading factors of 0 (yellow solid evolution track, equivalent to no outflow), 0.5 (pink dashed evolution track), 1 (teal dot-dashed evolution track) and 2 (red dotted evolution track). For each evolutionary track, the thick opaque line shows the evolutionary track for the properties of REBELS-25 measured in this paper and the two thin semi-transparent lines indicate the region bounded by the 16th and 84th percentiles of the SFR and stellar mass distributions from a sample of evolutionary tracks with starting conditions randomly selected from the uncertainty distributions of these parameters. The SFRs and stellar masses of observed high-z quiescent galaxies are indicated on the plot as open squares \citep{Valentino2020}, open diamonds \citep[the robust spectroscopic sample of][]{Schreiber2018a}, open pentagons \citep[the uncertain spectroscopic sample of][]{Schreiber2018a} and open hexagons  \citep[the photometric sample of][]{Schreiber2018a}.  }
	\label{fig:evolution_plot}

\end{figure*}

\begin{figure*}
	\centering
	\includegraphics[width=0.95\textwidth]{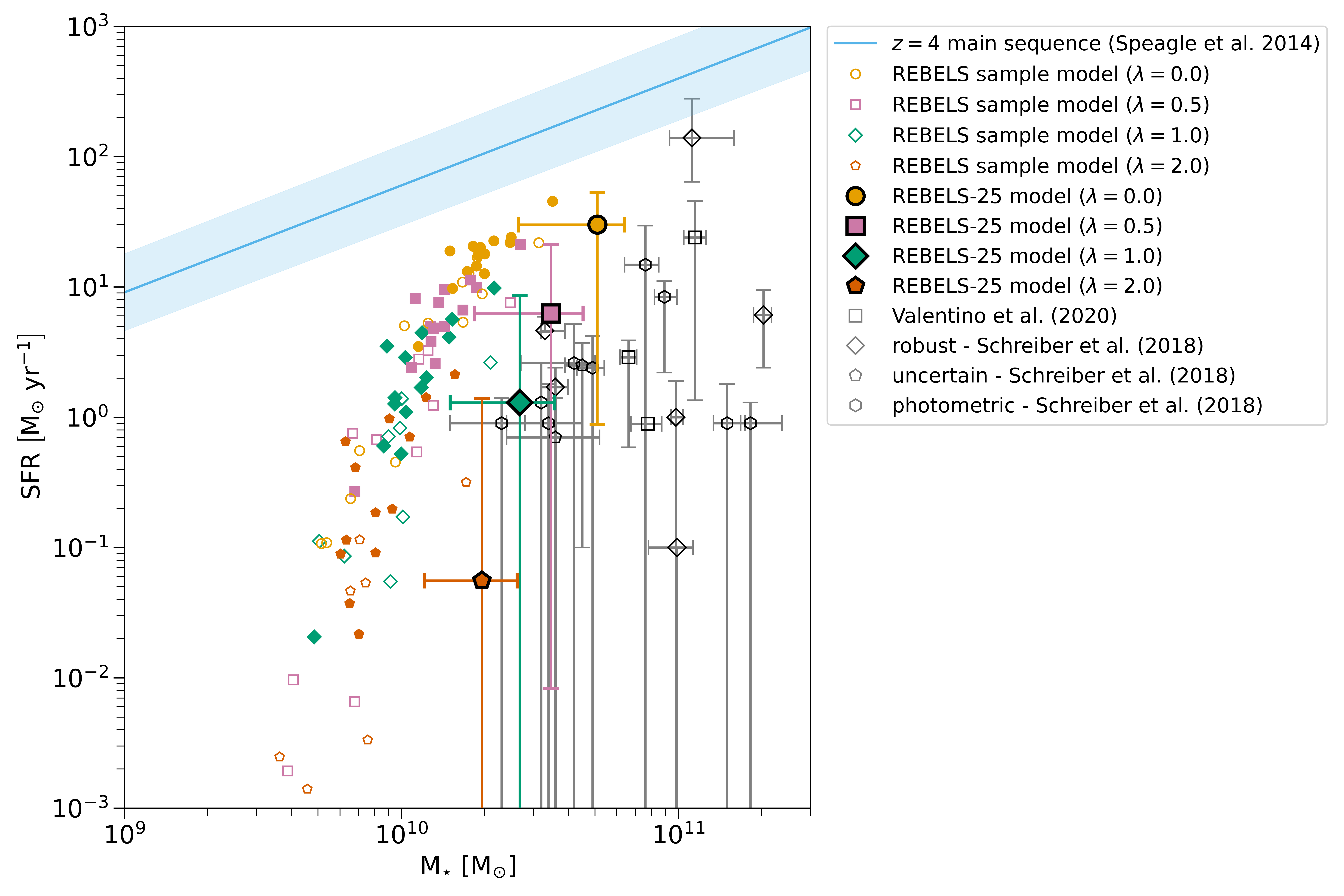}
	\caption{The predicted SFR against the stellar mass for the REBELS sample at $z=4$ derived from a simple conservative model under the assumption of continuous star formation with constant star formation efficiency and no inflow of gas. The REBELS galaxies are shown with filled markers where a continuum flux is reported by \citep{Inami2022} or with open markers where only an upper limit on the continuum flux (and thus only an upper limit on  the obscured star formation) is reported. The different marker  shape-colour combinations indicate the mass loading factor of the outflow model, either 0, which is equivalent to no outflow (yellow circles); 0.5 (pink squares); 1 (teal diamonds) or 2 (red pentagons). REBELS-25 is displayed with an additional solid black outline around the marker and with 1 $\sigma$ errorbars. The SFRs and stellar masses of quiescent galaxies observed at $z\sim4$ are indicated on the plot as open squares \citep{Valentino2020}, open diamonds \citep[the robust spectroscopic sample of][]{Schreiber2018a}, open pentagons \citep[the uncertain spectroscopic sample of][]{Schreiber2018a} and open hexagons \citep[the photometric sample of][]{Schreiber2018a}. The solid blue line shows the $z=4$ main sequence of \citet{Speagle2014} and the blue shaded area shows its associated 1-$\sigma$ error region.}
	\label{fig:sfr_mstar_plot}

\end{figure*}

We consider a simple evolutionary model that includes a declining SFR and the possibility of an outflow. Although inflows are thought to be important at the high redshifts of the EoR \citep[see e.g][]{SanchezAlmeida2014}, the assumption of no inflow, means that this model provides a conservative estimate of the stellar mass evolution of REBELS-25. An inflow would serve to increase the available gas mass for star formation, which would have the effect of increasing the SFR of the galaxy at a given redshift and thus its given stellar mass. Indeed observations of high-z quiescent galaxies \citep{Forrest2020}  and simulations \citep{Pallottini2017,Pallottini2022} suggest that these galaxies may have an SFH consistent with a sustained period of high SFR followed by late quenching. However, restricting ourselves to a conservative lower estimate of the stellar mass allows us to explore the possibility that REBELS-25 is a plausible progenitor of high-z quiescent galaxies.

We model the evolution of the galaxy forward in time, $t$, starting from its currently observed properties at t=0 (corresponding to its observed redshift at $z=\zciitwodp$). In this model, we assume an exponentially declining SFR, with fixed star formation efficiency,  $\varepsilon$, set by the observed SFR and molecular gas mass at  $t = 0$, i.e.
\begin{equation}
		\varepsilon = \dot{M_{*}}(0)/ M_{{\rm H}_{2}} (0).
\end{equation}
\noindent
where $\dot{M_{*}}(0) = \sfrtot{}  ~ \msol \yr^{-1}$  is our measurement of the star formation rate of REBELS-25 and $ M_{{\rm H}_{2}} (0) = \mmolcii{} ~ \msol $ is our measurement of the molecular gas mass. For the outflow, we  adopt a model where the mass outflow $\dot{M}_{\rm out}$ rate at $t$  is proportional to the SFR
\begin{equation}
	\dot{M}_{\rm out}  (t) =  \lambda \dot{M}_{*}(t) ,
\end{equation}
\noindent
where $\lambda$ is the outflow mass loading factor. We consider four scenarios for the strength of the outflow: $\lambda =0, 0.5, 1.0$ and $2.0$. We note that $\lambda =0$ represents the case with no outflow. With these assumptions the SFR at $t$, $\dot{M_{*}}  (t)$, is given by
\begin{equation}
	\dot{M}_{*}  (t) = \dot{M}_{*}(0)  e^{-\varepsilon\left[\left(1 -  \rho \right)+  \lambda \right]t } ,
\end{equation}
\noindent
where $ \rho =0.4$\footnote{A return fraction of  $\sim0.4$ is commonly adopted for the return fraction from a \citet{Chabrier2003} IMF \citep[see e.g.][]{MadauDickinson2014}, however, this value varies depending on the IMF adopted, properties of the ISM and assumptions one makes about stellar evolution \citep[see e.g.][]{Vincenzo2016}.}  is the fraction of gas returned to the interstellar medium via stellar feedback (primarily supernovae and stellar winds).

We randomly sample the uncertainty distribution of the currently observed properties of the galaxy in  order to determine the uncertainty on these evolutionary tracks. Specifically we create 10000 sets of initial properties, which are randomly sampled from independent Gaussian uncertainty distributions defined by the measured values of these parameters and their associated uncertainties. To determine the upper and lower uncertainty region for the evolutionary tracks, we calculate the 16th and 84th percentiles of the values of SFR and stellar mass from the evolutionary tracks generated from these randomly sampled starting conditions.

We display the resultant evolutionary tracks of SFR and stellar mass for REBELS-25 in Figure~\ref{fig:evolution_plot} compared against observed high-z quiescent galaxies from the literature. These tracks show that, under the assumptions of our model, REBELS-25 could evolve into a massive, low star-formation rate galaxy with properties consistent with the properties of a number of observed high-z quiescent galaxies. However, REBELS-25 does not have sufficient molecular gas mass currently to evolve into the most massive observed high-z quiescent galaxies. In order to do so REBELS-25 would require additional molecular gas mass such as through inflow or the addition of more stellar mass through mergers. We repeat this analysis for all other the galaxies in the REBELS sample with a spectroscopic redshift \citep[][Schouws et al. in prep.]{Bouwens2022}. We show the predicted SFR and stellar mass at $z\sim4$ for these galaxies compared to REBELS-25 and a number of high-z quiescent galaxies from the literature observed at $z\sim4$ in Figure~\ref{fig:sfr_mstar_plot}. Compared to the REBELS sample as a whole, REBELS-25 has the largest predicted stellar mass at $z=4$ regardless of the outflow mass loading factor that we adopt. 

REBELS-25 is able to reach a mass comparable to the lower end of the mass range of high-z quiescent galaxies identified by \citet{Glazebrook2017}, \citet{Schreiber2018} and  \citet{Valentino2020}, under the conditions of our model. However, the question arises as to how much gas inflow would be required for REBELS-25 to reach a mass similar to the more massive high-z quiescent galaxies that have been detected. We therefore return to consider inflows that we have so far not included in our model, by adding a simple constant inflow term. We compare the results of this model to the masses of the high-redshift quiescent galaxies identified by \citet{Glazebrook2017}, \citet{Schreiber2018} and  \citet{Valentino2020} with $z\geq 3.3$. In order for REBELS-25 to reach the median stellar mass of these galaxies at $z\sim4$, a constant inflow rate of $\sim25~\msol \yr^{-1}$ would be required, and to reach the maximum stellar mass a constant inflow of $\sim1000~\msol \yr^{-1}$ would be required.

\section{Conclusions}
\label{sec:conclusions}

We have presented ALMA \cii{} and dust continuum observations, obtained as part of the REBELS survey \citep{Bouwens2022}, of REBELS-25. REBELS-25 is a dusty ULIRG, with $L_{\rm IR} $   =  \lir{} \lsol{}, as determined from a monochromatic $\sim 158 $~\textmu m flux of 	\fluxcont~\textmu Jy	\citep{Inami2022}.  It has been confirmed at a spectroscopic redshift of \zciitwodp. It has a significant stellar mass  ($M_{*}$  =  \mstar; \citealt[][Stefanon et al. in prep.]{Bouwens2022}). In this paper we have used these ALMA observations in conjunction with HST observations of the UV to characterise REBELS-25. Our main conclusions are as follows:

\begin{itemize}

	\item We determine a \cii{} luminosity of $L_{\cii{}} = \lciival$, for the main double-peaked \cii{} emission component of REBELS-25. 
	\item We use our measurement of $L_{\cii{}}~\lsol{}$ to infer a significant molecular gas reservoir of \mmolcii ~\msol. Combined with our SFR measurement, this gives a short depletion time of $ t_{\rm depl, {\rm H}_{2}} = \tdepl{} $~Gyr, about half of the predicted main-sequence value.
	\item We determine a UV + IR star formation rate of  \sfrtot{}~\msol{}~$\yr^{-1}$, which places REBELS-25 a factor of four above an extrapolated \citet{Speagle2014} star formation main-sequence with our adopted stellar mass, which would classify it as starbursting galaxy. However, adopting the non-parametric stellar mass presented by \citet{Topping2022} or the main-sequence derived from the REBELS sample also presented by  \citet{Topping2022} would result in its classification as a main-sequence galaxy. In contrast, adopting the $z\sim5.5$ main sequence presented by \citet{Khusanova2021}  would lead to an even stronger classification of the galaxy as a starburst galaxy.
	\item The majority of the star formation in REBELS-25 is obscured star formation traced by IR emission (\sfrir{}~\msol{}~$\yr^{-1}$) and there is a small contribution from unobscured star formation as traced by UV (\sfruv{}~\msol{}~$\yr^{-1}$).
	\item We find that REBELS-25 exhibits a kinematic profile that appears consistent with a rotating disc. However, due to the large beam of our current \cii{} data, this could equally be the result of merging galaxies whose kinematics are smeared by the beam.
	\item We find, consistent with an earlier study by \citet{Schouws2022a}, that the dust continuum emission is offset from the UV clumps identified by \citet{Stefanon2019}. As with similar offsets in lower-redshift dusty star forming galaxies, this can be interpreted as resulting from UV clumps embedded in a disc and visible in areas of low-extinction or as resulting from a disturbed morphology due to a merger.
	\item We observe an additional \cii{} emission component that may indicate a minor merger or the presence of an outflow. Modelling this emission as coming from a merging galaxy, we measure a molecular gas mass  of \mmolmerger ~ \msol{} (about 18  per cent of REBELS-25's molecular gas reservoir). If we instead model the emission as coming from an outflow, we estimate a projected (deprojected) atomic mass outflow rate of  $\gtrsim 120$ \msol{} $\yr^{-1}$ ($\gtrsim \moutdproj{}$~\msol{}~$\yr^{-1}$), corresponding to atomic outflow mass loading factors of  $\gtrsim 0.6$ ( $\gtrsim 1.0$). However, due to the difficulty of determining the size of the outflow these numbers are quite uncertain.
	\item We also investigated the potential evolution of REBELS-25 by considering a simple, conservative evolutionary model with a star formation rate set by a constant star formation efficiency, no inflow of gas and the possibility of constant mass-loading factor outflows. From this model, we find that REBELS-25 could potentially evolve into a galaxy with properties consistent with the population of high-z quiescent galaxies observed at $z\sim4$ \citep[see e.g][]{Schreiber2018a,Valentino2020} without needing to acquire more molecular gas.

\end{itemize}

In summary, we find that REBELS-25 has significant existing stellar mass, which coupled with a large molecular gas reservoir and significant star formation rate make it a realistic progenitor of high redshift quiescent galaxies. In addition, it has the signatures of complex morphology including potential disc rotation and the possibility of merger and/or outflow activity that requires further high-resolution follow-up to confirm.

\section*{Acknowledgements}

We thank the anonymous referee for their detailed reading of the paper and useful comments that helped us to improve this paper. We wish to thank Yana Khusanova for making the intercept of the main sequence presented in \citep{Khusanova2021} available to us via private communication. This paper makes use of the following ALMA data: ADS/JAO.ALMA\#2019.1.01634.L and ADS/JAO.ALMA\#2017.1.01217.S. ALMA is a partnership of ESO (representing its member states), NSF (USA) and NINS (Japan), together with NRC (Canada), MOST and ASIAA (Taiwan), and KASI (Republic of Korea), in cooperation with the Republic of Chile. The Joint ALMA Observatory is operated by ESO, AUI/NRAO and NAOJ. This research is based on observations made with the NASA/ESA Hubble Space Telescope obtained from the Space Telescope Science Institute, which is operated by the Association of Universities for Research in Astronomy, Inc., under NASA contract NAS 5-26555. This paper made use of the following software packages: {\sc astrometry} \citep{Wenzl2022}, {\sc Astropy} \citep{AstropyCollaboration2013, AstropyCollaboration2018, AstropyCollaboration2022}, {\sc CASA} \citep{McMullin2007},  {\sc Matplotlib} \citep{Hunter2007}, {\sc NumPy}  \citep{Harris2020}, {\sc Pandas} \citep{McKinney2010}, {\sc Regions}  \citep{Bradley2022},  {\sc SciPy}  \citep{Virtanen2020} and {\sc spectral-cube}  \citep{Ginsburg2019}. APSH is part of Allegro, the European ALMA Regional Centre node in The Netherlands. Allegro is funded by The Netherlands Organisation for Scientific Research (NWO). JH gratefully acknowledges support of the VIDI research program with project number 639.042.611, which is (partly) financed by the Netherlands Organisation for Scientific Research (NWO). E.d.C. gratefully acknowledges the Australian Research Council as the recipient of a Future Fellowship (project FT150100079) and the ARC Centre of Excellence for All Sky Astrophysics in 3 Dimensions (ASTRO 3D; project CE170100013) M. R. is supported by the NWO Veni project "Under the lens" (VI.Veni.202.225). HI and HSBA acknowledge support from the NAOJ ALMA Scientific Research Grant Code 2021-19A. HI acknowledges support from JSPS KAKENHI Grant Number JP19K23462. RB acknowledges support from an STFC Ernest Rutherford Fellowship [grant number ST/T003596/1]. RE acknowledges funding from NASA JWST/NIRCam contract to the University of Arizona, NAS5-02015. MA acknowledges support from FONDECYT grant 1211951, ANID+PCI+INSTITUTO MAX PLANCK DE ASTRONOMIA MPG 190030 and ANID+PCI+REDES 190194 and ANID BASAL project FB210003. PD acknowledges support from the European Research Council's starting grant ERC StG-717001 (``DELPHI"), from the NWO grant 016.VIDI.189.162 (``ODIN") and the European Commission's and University of Groningen's CO-FUND Rosalind Franklin program. YF acknowledge support from NAOJ ALMA Scientific Research Grant number 2020-16B. IDL acknowledges support from ERC starting grant 851622 DustOrigin. AP acknowledges support from the ERC Advanced Grant INTERSTELLAR H2020/740120.  The Cosmic Dawn Center (DAWN) is funded by the Danish National Research Foundation under grant No. 140. This work has made use of data from the European Space Agency (ESA) mission {\it Gaia} (\url{https://www.cosmos.esa.int/gaia}), processed by the {\it Gaia} Data Processing and Analysis Consortium (DPAC, \url{https://www.cosmos.esa.int/web/gaia/dpac/consortium}). Funding for the DPAC has been provided by national institutions, in particular the institutions participating in the {\it Gaia} Multilateral Agreement.

\section*{Data Availability}

The ALMA observations used in this article are available in the ALMA archive \url{https://almascience.eso.org/aq/} and can be accessed with their project codes: 2019.1.01634.L and 2017.1.01217.S. The COSMOS-DASH HST image used  available from the Mikulski Archive for Space Telescopes on the COSMOS-DASH page:  \url{https://archive.stsci.edu/hlsp/cosmos-dash}. The Gaia DR3 data underlying this paper is available from the GAIA archive \url{https://gea.esac.esa.int/archive/}. The data underlying this article will be shared on reasonable request to the corresponding author.

\bibliographystyle{mnras}
\bibliography{paper.bib} %

\appendix

\section{Models fitted to the spectrum}
\label{sec:spectral_models}

In Section~\ref{sec:spectral_fitting}, we fit models to the spectrum of REBELS-25 (see Figure~\ref{fig:cii_spectrum}), which are made up of a linear sum of up to three Gaussian functions of the form

\begin{equation}
	\label{eq:gaussian_model}
  	c  \exp \left(\frac{- (x - \mu)^{2} }{2 \sigma^{2}} \right) .
\end{equation}

\noindent
The models have either one or two Gaussians to fit the main \cii{} component and either one or no Gaussians to fit the secondary \cii{} component. The model of the secondary \cii{} component is either a broad Gaussian to represent an outflow or a narrow Gaussian to represent a merging galaxy. We also impose constraints on the parameters of our fits. For the Gaussian(s) representing the main component we require \mbox{$ c > 0$ mJy }and \mbox{$ -350 \kms{} < \mu < 250 \kms{}$} . For the Gaussian representing a merger, we require \mbox{$ c > 0$ mJy }and \mbox{$ 250 \kms{} < \mu < 750 \kms{}$.} Lastly, for the Gaussian representing an outflow we require \mbox{$ c > 0$} mJy, \mbox{$ -350 \kms{} < \mu < 750 \kms{}$} and \mbox{$ \sigma > 150$ \kms{} }. The best-fitting parameters and the reduced $\chi^2$ values of the models that we fit are displayed in Table~\ref{tab:spectrum_models}.

\begin{table*}
	\centering\setcellgapes{2pt}\makegapedcells
	\caption{A summary of the models fitted to the spectrum in Section~\ref{sec:spectral_fitting}. The models are a linear sum of between one and three Gaussians of the form in Equation~\ref{eq:gaussian_model}.}
	\label{tab:spectrum_models}
	\begin{tabular}{lcccccccccc}
		\hline
		Model Name & $\chi^{2}_{\rm red} $ & $c_{1}$ & $\mu_{1}$ & $\sigma_{1}$ & $c_{2}$ &  $\mu_{2}$ & $\sigma_{2}$ & $c_{3}$ &  $\mu_{3}$ & $\sigma_{3}$ \\
		$\mathrm{}$ & $\mathrm{}$ & $\mathrm{mJy}$ & $\mathrm{km\,s^{-1}}$ & $\mathrm{km\,s^{-1}}$ & $\mathrm{mJy}$ & $\mathrm{km\,s^{-1}}$ & $\mathrm{km\,s^{-1}}$ & $\mathrm{mJy}$ & $\mathrm{km\,s^{-1}}$ & $\mathrm{km\,s^{-1}}$  \\
		
 		\hline
		Double Gaussian + merger & 1.51 & $4.2 \pm 0.2$ & $-43 \pm 10$ & $91 \pm 9$ & $2.8 \pm 0.4$ & $134 \pm 9$ & $51 \pm 8$ & $0.9 \pm 0.2$ & $514 \pm 22$ & $102 \pm 22$ \\
		Double Gaussian + outflow & 1.64 & $3.8 \pm 0.3$ & $-45 \pm 9$ & $77 \pm 9$ & $2.6 \pm 0.3$ & $124 \pm 9$ & $48 \pm 8$ & $0.6 \pm 0.2$ & $97 \pm 79$ & $378 \pm 90$ \\
		Single Gaussian + merger & 1.72 & $4.2 \pm 0.2$ & $4 \pm 6$ & $133 \pm 6$ & $0.9 \pm 0.2$ & $531 \pm 21$ & $90 \pm 22$ &  &  &  \\
		Single Gaussian + outflow & 1.92 & $3.7 \pm 0.3$ & $-2 \pm 8$ & $119 \pm 11$ & $0.5 \pm 0.3$ & $172 \pm 131$ & $342 \pm 96$ &  &  &  \\
		Double Gaussian & 1.95 & $3.7 \pm 0.3$ & $20 \pm 11$ & $137 \pm 8$ & $1.3 \pm 0.4$ & $-71 \pm 12$ & $37 \pm 15$ &  &  &  \\
		Single Gaussian & 2.02 & $4.2 \pm 0.2$ & $5 \pm 6$ & $134 \pm 6$ &  &  &  &  &  &  \\
	\end{tabular}
\end{table*}

\section{Significance of the secondary \cii{} emission component}
\label{sec:appendix_data_split}

\begin{figure*}
	\centering
	\includegraphics[width=0.9\textwidth]{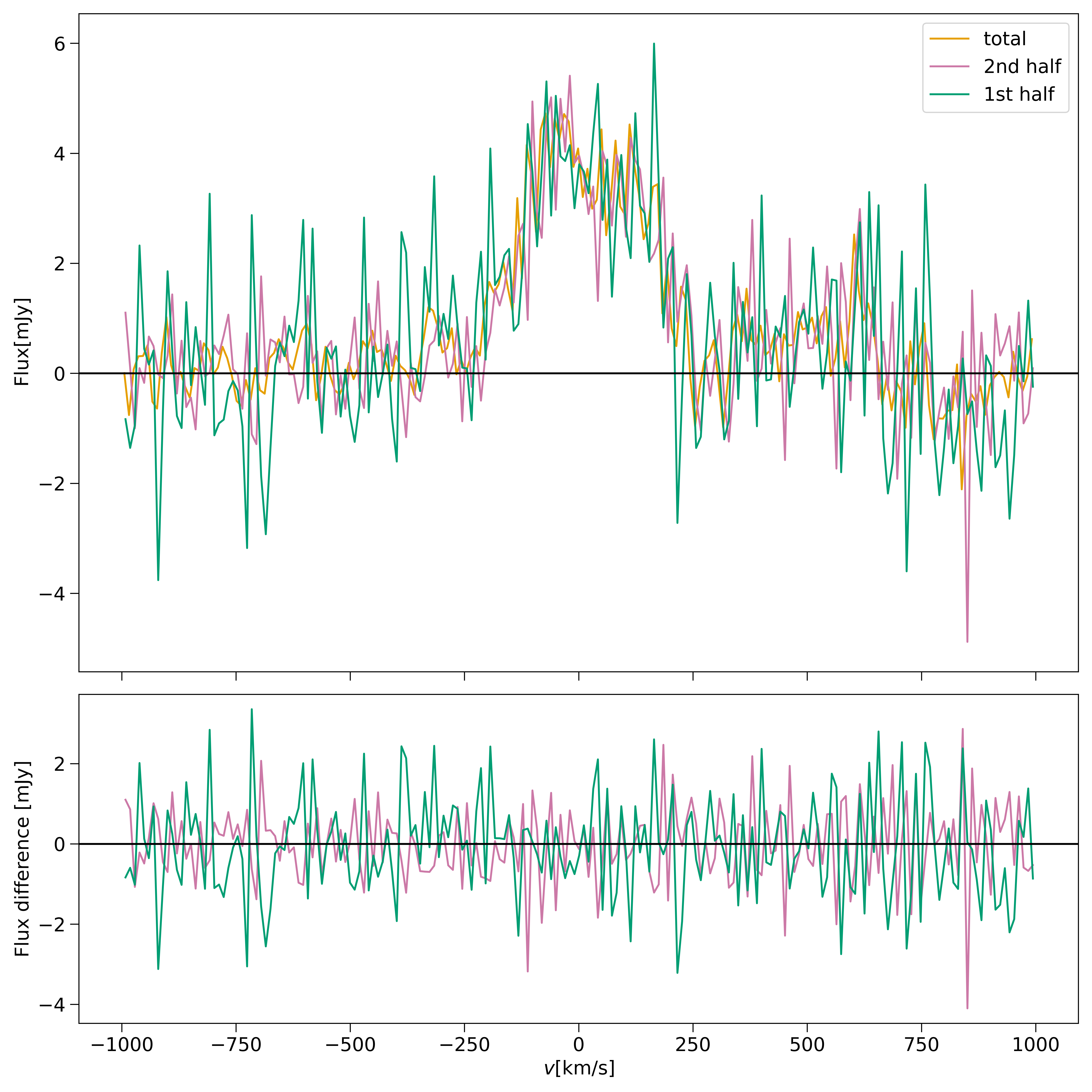}
	\caption{ [Top panel] the spectra extracted from data cubes created with all the ALMA data (yellow line),  the first half of the ALMA data (pink line)  and the second half of the ALMA data (teal line).[ Bottom panel] the flux difference between the two half-data spectra and the total spectrum. Both the main and secondary \cii{} components remain visible in both of the half-data spectra.}
	\label{fig:spectrum_split}

\end{figure*}

We identify both a bright (double-peaked) main \cii{} component and a fainter secondary \cii{} component. In order to investigate the significance of the secondary \cii{} component, we split the data in half and reimage it. The first half of the data was observed on the 22nd of November 2019 and the second half of the data was observed on the 25th of November and the 1st of December 2019. The resulting spectra can be seen in Figure~\ref{fig:spectrum_split}  in comparison to the spectrum produced from all of the data combined. We then collapse the datacubes over the line  to create a moment-0 map and measure the peak signal to noise in the image plane for the secondary  component.  With both haves of the data together, we measure a peak signal-to-noise ratio of \offsetsnr{}. We see by inspecting the two half-data spectra  that the secondary  \cii{} component remains visible in both halves of the data. We determine a peak signal-to-noise ratio of 4.0 and 3.1 for the secondary component in first and second halves of the data. We thus conclude that the secondary  component is significant.

\bigskip
\noindent
{\small\textit{%
\noindent
$^{1}$ Leiden Observatory, Leiden University, P.O. Box 9513, 2300 RA Leiden, The Netherlands
\\
$^{2}$ International Centre for Radio Astronomy Research, University of Western Australia, 35 Stirling Hwy, Crawley, 26WA 6009, Australia
\\
$^{3}$ ARC Centre of Excellence for All Sky Astrophysics in 3 Dimensions (ASTRO 3D), 2601, Australia
\\
$^{4}$ Faculty of Electrical Engineering, Mathematics and Computer Science, Delft University of Technology, Delft, The Netherlands
\\
$^{5}$ Hiroshima Astrophysical Science Center, Hiroshima University, 1-3-1 Kagamiyama, Higashi-Hiroshima, Hiroshima 739-8526, Japan
\\
$^{6}$ Departament d'Astronomia i Astrof\'isica, Universitat de Val\`encia, C. Dr. Moliner 50, E-46100 Burjassot, Val\`encia,  Spain
\\
$^{7}$ Dipartimento di Fisica, Sapienza, Universit\`{a} di Roma, Piazzale Aldo Moro 5, I-00185 Roma, Italy
\\
$^{8}$ INAF/Osservatorio Astrofisico di Arcetri, Largo E. Fermi 5, I-50125 Firenze, Italy
\\
$^{9}$ Sapienza School for Advanced Studies, Viale Regina Elena 291, I-00161 Roma Italy
\\
$^{10}$ INAF/Osservatorio Astronomico di Roma, via Frascati 33, I-00078 Monte Porzio Catone, Roma, Italy
\\
$^{11}$ Istituto Nazionale di Fisica Nucleare, Sezione di Roma1, Piazzale Aldo Moro 2, I-00185 Roma Italy
\\
$^{12}$ Kapteyn Astronomical Institute, University of Groningen, NL-9700 AV
Groningen, the Netherlands
\\
$^{13}$ Astrophysics Research Institute, Liverpool John Moores University, 146 Brownlow Hill, Liverpool L3 5RF, UK
\\
$^{14}$ Jodrell Bank Centre for Astrophysics, Department of Physics and Astronomy, School of Natural Sciences, The University of Manchester,Manchester, M13 9PL, UK
\\
$^{15}$ Steward Observatory, University of Arizona, 933 N Cherry Ave, Tucson, AZ 85721, USA
\\
$^{16}$ Departmento de Astronomia, Universidad de Chile, Casilla 36-D, Santiago 7591245, Chile
\\
$^{17}$ Centro de Astrofisica y Tecnologias Afines (CATA), Camino del Observatorio 1515, Las Condes, Santiago, 7591245, Chile
\\
$^{18}$ Observatoire de Gen\`{e}ve, CH-1290 Versoix, Switzerland
\\
$^{19}$ Cosmic Dawn Center (DAWN), Niels Bohr Institute, University of Copenhagen, Jagtvej 128, København N, DK-2200, Denmark
\\
$^{20}$National Astronomical Observatory of Japan, 2-21-1, Osawa, Mitaka, Tokyo, Japan 
\\
$^{21}$ Nucleo de Astronomia, Facultad de Ingenieria y Ciencias, Universidad Diego Portales, Av. Ejercito 441, Santiago, Chile \\
$^{22}$ Scuola Normale Superiore, Piazza dei Cavalieri 7, I- 50126 Pisa, Italy \\
$^{23}$ Waseda Research Institute for Science and Engineering, Faculty of Science and Engineering, Waseda University, 3-4-1 Okubo, Shinjuku, Tokyo 169-8555, Japan \\
$^{24}$ Sterrenkundig Observatorium, Ghent University, Krijgslaan 281-S9, B-
9000 Gent, Belgium \\
$^{25}$ Dept. of Physics \& Astronomy, University College London, Gower Street, London WC1E 6BT, UK  \\
$^{26}$ Centre for Astrophysics and Supercomputing, Swinburne University of Technology, PO Box 218, Hawthorn, VIC 3112, Australia \\ 
$^{27}$ I. Physikalisches Institut, Universit\"at zu K\"oln, Z\"ulpicher Strasse 77, D-50937 K\"oln, Germany
}}

\bsp	%
\label{lastpage}
\end{document}